\documentclass[aps,prb,twocolumn,amssymb,floatfix,epsfig]{revtex4}
\usepackage{graphicx,epsfig,psfrag,subfigure,amsopn,epstopdf}
\usepackage{color,ulem,verbatim,bbold,float,wrapfig}
\usepackage{hyperref}
\usepackage{hypcap}
\usepackage{amssymb,amsbsy,amsmath,mathrsfs}
\newcommand\beq{\begin{equation}}
\newcommand\eeq{\end{equation}}
\newcommand\bea{\begin{eqnarray}}
\newcommand\eea{\end{eqnarray}}
\newcommand\bsq{\begin{subequations}}
\newcommand\esq{\end{subequations}}

\newcommand\non{\nonumber}

\newcommand\bib{\bibitem}

\newcommand\al{\alpha}
\newcommand\be{\beta}
\newcommand\ga{\gamma}
\newcommand\de{\delta}
\newcommand\De{\Delta}
\newcommand\ep{\epsilon}
\newcommand\si{\sigma}
\newcommand\ka{\kappa}

\newcommand\pa{\partial}
\newcommand\la{\langle}
\newcommand\ra{\rangle}

\newcommand\ta{\theta}
\newcommand\dg{\dagger}

\newcommand\vk{{\vec k}}
\newcommand\vn{{\vec n}}
\newcommand{\hn}{\hat{n}}
\newcommand\clyM{\mathcal{M}}
\newcommand\red{\textcolor{red}}
\begin{document}

\title{Topological magnons in a kagome lattice spin system with $XXZ$
and Dzyaloshinskii-Moriya interactions}

\author{Ranjani Seshadri and Diptiman Sen}

\affiliation{\small{Centre for High Energy Physics, Indian Institute of 
Science, Bengaluru 560012, India}} 

\date{\today}

\begin{abstract}
We study the phases of a spin system on the kagome lattice with 
nearest-neighbor $XXZ$ interactions with anisotropy ratio $\De$ and 
Dzyaloshinskii-Moriya interactions with strength $D$. In the classical 
limit where the spin $S$ at each site is very large, we find a rich phase 
diagram of the ground state as a function of $\De$ and $D$. There are five 
distinct phases which correspond to different ground-state spin configurations 
in the classical limit. We use spin-wave theory to find the bulk energy bands
of the magnons in some of these phases. We also study a strip of the system 
which has infinite length and finite width; we find states which are localized 
near one of the edges of the strip with energies which lie in the gaps of the 
bulk states. In the ferromagnetic phase in which all the spins point along the 
$+ \hat z$ or $- \hat z$ direction, the bulk bands are separated from each 
other by finite energy gaps. This makes it possible to calculate the Berry 
curvature at all momenta, and hence the Chern numbers for every band; the
number of edge states is related to the Chern numbers. Interestingly, we 
find that there are four different regions in this phase where the Chern 
numbers are different. Hence there are four distinct topological phases even 
though the ground-state spin configuration is identical in all these phases. 
We calculate the thermal Hall conductivity of the magnons as a function of the 
temperature in the above ferromagnetic phase; we find that this can distinguish 
between the various topological phases. These results are valid for all values 
of $S$. In the other phases, there are no gaps between the different bands; 
hence the edge states are not topologically protected.
\end{abstract}

\maketitle

\section{Introduction}
\label{sec:intro}

For the last several years topological phases of matter have been studied
extensively in a variety of systems both theoretically and 
experimentally~\cite{qi,khan,mous}. A hallmark of such systems is that the 
energy bands of the bulk states have gaps, and there are states which lie at 
the edges of the system and whose energies lie in the gaps of the bulk 
bands. In addition, these systems have a bulk-boundary correspondence; the 
bulk bands are characterized by some topological invariants which are 
integers, and these give the number of edge states with a given momentum.

A relatively new entrant to this field is a class of spin systems in 
which the magnons (spin waves) have a topologically nontrivial nature. These 
topological magnons have been studied theoretically in ferromagnets on a kagome
lattice~\cite{lifa,mook1} and in spin systems with magnetic dipolar 
interactions~\cite{shind}. Observations from neutron scattering experiments 
have been reported supporting some of the theoretical analysis of magnons in 
such systems~\cite{chis}. Topological magnons have been studied 
on a honeycomb lattice as well~\cite{ower1,ower2,mish,pers}. The 
topologically nontrivial nature of magnons leads to robust edge 
states ~\cite{mook2,pant} and a thermal Hall conductivity which has been 
investigated both theoretically~\cite{lee,naka1,laur} and 
experimentally~\cite{hir}. There is also experimental evidence for the magnon 
thermal Hall effect in a pyrochlore lattice ferromagnet~\cite{onos}.
Topological magnons have also been studied in antiferromagnetic 
systems~\cite{rom,okum,naka2}. While many of the spin systems known to 
possess topological magnons have Dzyaloshinskii-Moriya interactions 
(DMIs)~\cite{dzya,mori,elha,tail},
topological systems without such interactions are also known~\cite{naka2}.
A necessary though not sufficient condition for a two-dimensional
system to have topological magnon bands with nonzero Chern numbers is that 
the Hamiltonian should break the symmetry under reflection about some axis,
thus giving a chirality to the system.

It has been shown in a variety of such systems that there are topological 
phases in which the bulk bands of the magnons are gapped with respect to each 
other, and they have nonzero values of some topological invariant which 
gives the number of edge states at the boundaries of finite-sized systems. 
The different topological phases typically have different ground states.

In this paper, we will study a system consisting of spins on a kagome lattice 
with anisotropic $XXZ$ and Dzyaloshinskii-Moriya interactions between 
nearest-neighbor spins. (While the effects of Dzyaloshinskii-Moriya 
interactions have been extensively studied earlier~\cite{elho}, the combined 
effects of $XXZ$ and Dzyaloshinskii-Moriya interactions have not been 
investigated as thoroughly; however, see Refs.~\onlinecite{ess1,ess2}). We 
will show that there are several topological phases of magnons all of which 
have the same ground state (a similar phenomenon is known to occur in 
other spin systems~\cite{mook1,laur}). We will show that the different 
topological phases can be distinguished from each other through the 
temperature dependence of the thermal Hall 
conductivity~\cite{kats,mats1,mats2}. These results turn out to be valid for 
any value of the spin. (For reviews of the thermal Hall effect of magnons, 
see Refs.~\onlinecite{han,mura}).

The plan of the paper is as follows. In Sec.~\ref{sec:model}, we introduce a
model of spins on a kagome lattice with nearest-neighbor $XXZ$ 
interactions with anisotropy ratio $\De$ and a Dzyaloshinskii-Moriya 
interactions with strength $D$. In the classical limit in which 
the spin $S \to \infty$, we find the ground states over a wide range of 
$\De$ and $D$. We discover that there are five different phases numbered $I$ 
to $V$ which are separated pairwise by phase transition lines. Of these, 
phase $I$ is the simplest; here the ground state consists of all spins 
pointing in the $+ \hat z$ or $- \hat z$ direction. In the ground state of 
phase $II$, all the spins point in the same direction which can be along any 
direction in the $x-y$ plane. In phases $III$ and $IV$, the three spins in 
each triangle of the kagome lattice form a configuration in which they lie in 
the $x-y$ plane at angles which differ from each other by $2\pi/3$ 
in a clockwise or anticlockwise sense. The ground state in phase $V$ has the 
most complex spin configuration. In Sec.~\ref{sec:phaseI}, we study linear spin
wave theory in phase $I$ by using the Holstein-Primakoff transformation and
expanding the Hamiltonian to the next order in a $1/S$ expansion. Since
the unit cell of the kagome lattice consists of three spins forming a 
triangle, spin-wave theory gives three bands of magnon energies. We find that
the bands are all separated from each other by gaps. This allows us to
calculate the Berry curvature in each band for all values of the momentum
in the Brillouin zone (BZ). Integrating the Berry curvature gives the Chern
numbers of the different bands. We then find that phase $I$ consists
of four distinct topological phases; these are separated from each other
by lines where the gaps between pairs of bands vanish. Next we consider a 
strip of the kagome lattice which is infinitely long and has a finite width. 
We discover that there are magnons which are localized near one of the two 
edges of the strip; the energies of these edge magnons lie in the gaps of the 
bulk bands. The number of edge magnons is closely related to the Chern 
numbers. We then calculate the thermal Hall conductivity and show that it 
has different values in the different topological phases. All these results 
turn out to be valid for all values of 
$S$. In Sec.~\ref{sec:phaseII}, we study spin-wave theory in phases $II$ and 
$III$. The three magnon bands are not separated from each other by gaps in 
these phases. Hence the Berry curvatures are ill-defined at certain momenta and
the Chern numbers cannot be calculated. Although edge states appear in these 
phases also, they are not topologically protected. In Sec.~\ref{sec:discussion}
we summarize our results and point out some directions for future studies.

\section{Spins on a kagome lattice with $XXZ$ and Dzyaloshinskii-Moriya 
interactions}
\label{sec:model}

\begin{figure}
\subfigure[~Kagome lattice]{\includegraphics[width=7.3cm]{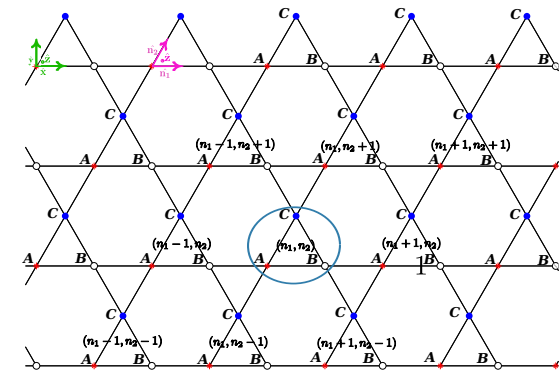} 
\label{lattice}} \\
\subfigure[~Phase diagram]{\includegraphics[width=7.3cm]{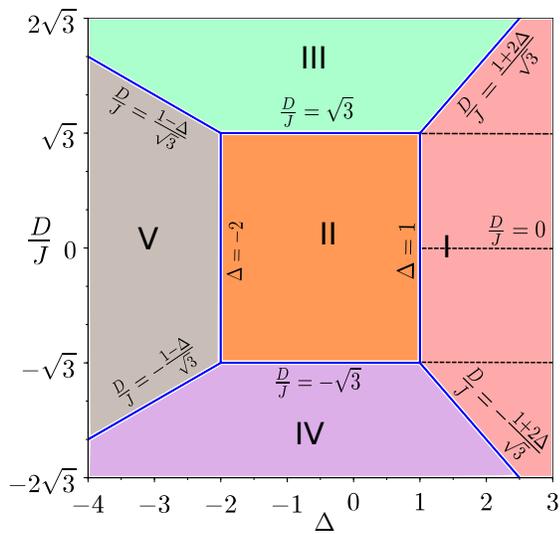}
\label{phase_diag}} \\
\subfigure[~Spin configurations]{\includegraphics[width=7.3cm]{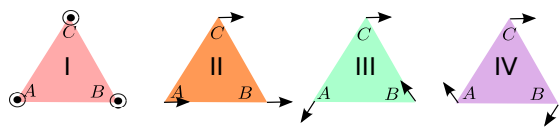}
\label{phase1_config}}
\caption{A kagome lattice can be viewed as a triangular lattice of unit cells
each of which is also a triangle with three sites labeled $A$, $B$ and $C$.
One of the unit cells is shown as a blue circle in (a). The geometry shown in 
(a) is used for our edge state calculations; thus the bottom edge 
is a straight edge whereas the top edge is jagged. The top left corner 
(in green) shows the coordinate axes. Adjacent to it (in magenta) is shown the 
nonorthogonal $n_1 - n_2$ coordinate axes used for our calculations. The 
$\hat z$-axis points out of the plane. Classically the system has five phases 
shown in different colors in (b); the corresponding spin orientations in the 
ground state are shown in (c). Although phase $I$ (red) is a single phase 
in terms of the ground-state spin configuration, it consists of four distinct 
topological phases which are separated from each other by the dotted 
horizontal lines $D/J = 0,\pm{\sqrt{3}}$.} \label{classical} \end{figure}

We consider a system of spins on a kagome lattice with $XXZ$ and 
Dzyaloshinskii-Moriya interactions (DMI) between nearest neighbors.
A schematic picture of the system is shown in Fig.~\ref{lattice}. The
unit cell of the kagome lattice is given by a triangle consisting
of sites $A$, $B$ and $C$ as shown. We will denote the coordinates of a unit 
cell by $\vn =(n_1,n_2)$, and the spin operators at the three sites by $A_\vn$,
$B_\vn$ and $C_\vn$ respectively. The Hamiltonian of the system is a sum of 
the $XXZ$ interactions ($H_{\De}$) and DMI ($H_{DM}$), where
\bea H &=& H_{\De} + H_{DM}, \label{Hamiltonian} \\
H_{\De} &=& - \displaystyle \sum_{\substack{\la \vn \vn' \ra \non \\ 
\al=x,y,z}} J_\al (A^{\al}_{\vn} B^{\al}_{\vn'}+B^{\al}_{\vn}C^{\al}_{\vn'}
+C^{\al}_{\vn}A^{\al}_{\vn'}), \non \\
H_{DM} &=& D ~\hat{z} \cdot\displaystyle \sum_{\la \vn \vn' \ra} 
(\vec{A}_{\vn} \times\vec{B}_{\vn'}+\vec{B}_{\vn}\times\vec{C}_{\vn'}+
\vec{C}_{\vn}\times\vec{A}_{\vn'}), \non \\ 
{\rm where} && J_\al ~=~ \begin{cases} 
~~~J & \text{if} ~~~ \al = x ~~{\rm or}~~ y, \\
\De J & \text{if}~~~ \al = z, \\ 
\end{cases} \non \eea
and $\la \vn \vn' \ra$ denotes nearest-neighbor pairs. We will
assume $J>0$ in this paper, i.e., the interaction for the $x-y$ components
of the spins is ferromagnetic. Note that we have only considered DMI
of the out-of-plane type (only $D_z \ne 0$); this is because of Moriya's 
symmetry rules which imply that only such terms can be present on the 
kagome lattice~\cite{mori}. (This, however, is true only for perfect kagome 
lattices, i.e., for lattices which are not locally tilted. For systems with 
lower symmetry, there can be in-plane components of the DMI, albeit very small
\cite{elha}). The DMI breaks the symmetry under reflection about the $x$-axis
and the $y$-axis. The Hamiltonian in Eq.~\eqref{Hamiltonian} has
two parameters, the anisotropy ratio $\De$ which is dimensionless and the DMI 
strength $D$ (the ratio $D/J$ is dimensionless, hence we will often use this 
parameter below). For $\De =1$, $H_{\De}$ describes an isotropic ferromagnet.
We note that both the terms in the Hamiltonian in Eq.~\eqref{Hamiltonian}
are invariant under arbitrary spin rotations in the $x-y$ plane, i.e., around 
the $\hat z$ axis. We will see below that this continuous symmetry has 
implications for the magnon energy spectrum in some of the phases.

The system described in Eq.~\eqref{Hamiltonian} becomes classical in the limit
that the spin $S \to \infty$, where $\vec{A}_\vn^2 = \vec{B}_\vn^2 = 
\vec{C}_\vn^2 = S(S+1) \hbar^2$. In this limit we can find the ground state
spin configuration as follows. We assume that in the ground state, all the 
unit cells (i.e., triangles) look identical. This means that the ground state 
of the full kagome lattice is the same as that of the three spins at the 
vertices of a triangle. Classically, we can write the components of 
$\vec{A}_\vn, \vec{B}_\vn$ and $\vec{C}_\vn$ in terms of the polar coordinates 
$\ta$ and $\phi$. Since the Hamiltonian is invariant under spin rotations in 
the $x-y$ plane, we can choose, say, $\phi_C =0$. We will therefore write 
\bea \vec{A}_\vn &=& S \big(\sin{\ta_A} \cos{\phi_A} , \sin{\ta_A} 
\sin{\phi_A}, \cos{\ta_A} \big), \non \\ 
\vec{B}_\vn &=& S \big(\sin{\ta_B} \cos{\phi_B} , \sin{\ta_B} 
\sin {\phi_B}, \cos{\ta_B} \big), \non \\
\vec{C}_\vn &=& S \big(\sin {\ta_C} , 0, \cos{\ta_C} \big), \label{taph} \eea
for all values of $\vn$. We thus have to consider five parameters, $p_i = 
\ta_A, \ta_B, \ta_C, \phi_A, \phi_B$. We then 
numerically find the spin configuration which minimizes the energy of the 
Hamiltonian given in Eq.~\eqref{Hamiltonian} for different values of the
five parameters. We find five phases as shown in Fig.~\ref{phase_diag}. In 
phase $I$, all the spins point along the $+ \hat z$ (or $- \hat z$) direction;
this is a ferromagnetic configuration along the $\hat z$ axis. In phase $II$,
all the spins points along the $\hat x$ direction (this direction is
fixed due to our choice $\phi_C = 0$); hence we have a 
ferromagnetic configuration along the $\hat x$ direction. Thus both
phases $I$ and $II$ have colinear spin configurations.
In phase $III$, the spins again lie in the $x-y$ plane, but they rotate by 
$2\pi/3$ in the clockwise direction as we go around the triangle from $A$ 
to $B$ to $C$ (see Fig.~\ref{phase1_config}. In phase $IV$, the spins 
lie in the $x-y$ plane and rotate by $2\pi/3$ in the anticlockwise direction 
as we go around the triangle. Thus phases $III$ and $IV$ have coplanar 
configurations. (We can see from Eq.~\eqref{Hamiltonian} that changing $D 
\to - D$ is equivalent to carrying out the unitary transformation $S_i^y 
\to - S_i^y, ~S_i^z \to - S_i^z$, and $S_i^x \to S_i^x$ on every site $i$, 
namely, rotating by $\pi$ around the $\hat x$ axis. We can see from 
Fig.~\ref{phase1_config} that this transformation changes the spin 
configuration in phase $III$ to that in phase $IV$. We will study only 
phase $III$ below since the analysis in phase $IV$ is similar). In phase 
$V$, we have found numerically that the spins do not generally lie in the 
same plane; this makes it difficult to study this phase analytically. 

In Fig.~\ref{phase_diag}, the phase transition lines lying between two of the 
four phases $I$, $II$, $III$ and $IV$ are lines on which the ground-state 
energies of the two appropriate phases are equal. We can find out as
follows if these lines correspond to a phase transition of first order or 
higher order. We calculate the first derivative of the ground-state energy 
along a line which crosses a phase transition line perpendicularly. If the 
derivative approaches different values as we move towards the phase transition 
line from the two sides, it corresponds to a first order transition; otherwise 
we call it a higher order transition. We find that all the phase transition 
lines between phases $I - IV$ are of first order. However, this method cannot 
be used to characterize the phase transition lines lying between phase $V$ and 
phases $II - IV$; since we do not have an analytical expression for the 
ground-state energy in phase $V$, we do not know what its first derivative is 
when we approach one of its phase transition lines.

Next, we can study the stability of phases $I ~-~ IV$. To do this, we slightly
perturb the five parameters $p_i$ away from their ground-state values. We then 
find the matrix of second derivatives of the ground-state energy; the entries 
of this matrix are given by $S_{ij} = \pa^2 E_0 /\pa p_i \pa p_j$. If all the 
eigenvalues of this matrix are positive, the phase is stable; if any one of 
the eigenvalues is negative, it is unstable. We find that the phases $I - IV$ 
are stable in the regions shown for them in Fig.~\ref{phase_diag}. This method 
also gives the phase transition lines between the phases $II ~-~ IV$ and 
phase $V$; those are the lines where one of the eigenvalues of $S_{ij}$ 
calculated in phases $II, III$ and $IV$ becomes zero. 

We will now present some examples of 
the kinds of calculations mentioned above. We first look at the 
neighborhood of the line $\De = 1$ in Fig.~\ref{phase_diag} which lies between 
phases $I$ and $II$. We consider a ferromagnetic spin configuration given by 
$\phi_A = \phi_B = \phi_C = 0$ and $\ta_A = \ta_B = \ta_C = \ta$. For this 
configuration, the energy of a single triangle is given by 
\beq E (\ta) ~=~ - 3 J S^2 (\De \cos^2 \ta + \sin^2 \ta). \label{Eta} \eeq
For $\De > 1$, $E (\ta)$ has a minimum value of $E_0 = - 3 J S^2 \De$ at $\ta 
= 0$ or $\pi$ which corresponds to all the spins pointing along the $+ \hat z$
or $- \hat z$ direction (phase $I$), while for $\De < 1$, $E(\ta)$ has a 
minimum value of $E_0 = - 3 J S^2$ at $\ta = \pi/2$ which corresponds to spins 
pointing along the $+ \hat x$ direction (phase $II$). The two ground-state 
energies match at $\De =1$, but $\pa E_0 /\pa \De$, which is equal to $-3 J 
S^2$ and zero in phases $I$ and $II$, do not match at $\De = 1$. Hence the 
phase transition on the line $\De = 1$ is of first order. To examine the 
stability of phase $I$, we note from Eq.~\eqref{Eta} that if $\ta$ is changed 
from zero to $\ep$ (where $\ep$ is small), the energy becomes $-3 J S^2 (\De - 
\De \ep^2 + \ep^2)$. Hence $\pa^2 E_0 /\pa \ep^2 = 6 J S^2 (\De - 1)$. This is 
positive and therefore phase $I$ is stable if $\De > 1$. To examine phase 
$II$, we change $\ta$ from $\pi/2$ to $\pi/2 + \ep$. The energy is then given 
by $-3 J S^2 (1 - \ep^2 + \De \ep^2)$. Hence $\pa^2 E_0 /\pa \ep^2 = 6 J S^2 
(1 - \De)$. Thus phase $II$ is stable if $\De < 1$. This gives us an 
understanding of the line $\De = 1$ which separates phases $I$ and $II$.

Next we consider the line $\De = -2$ which separates phases $II$ and $V$
(see Fig.~\ref{phase_diag}).
It turns out that phase $V$ is difficult to study analytically (except 
on the line $D=0$ which will be discussed in the next paragraph). We will
therefore only consider what happens if we approach $\De = -2$ from phase 
$II$, namely, from the region $\De > -2$. Let us consider a spin configuration
given by $\phi_A = \phi_B = \phi_C = 0$, $\ta_A = \pi/2 +\al$, $\ta_B = \pi/2 
+\be$, and $\ta_C = \pi/2 + \ga$. We then find that the energy of a single 
triangle is given by
\bea E (\al, \be, \ga) &=& - 3JS^2 ~+~ JS^2 ~(\al^2 + \be^2 + \ga^2) \non \\
&& -~ J S^2 \De ~(\al \be + \be \ga + \ga \al) \label{en1} \eea
up to second order in $\al, \be, \ga$. We can write the last two terms
in Eq.~\eqref{en1} in terms of a matrix $M$ as $(JS^2 /2) ~(\al, \be, \ga) ~M~
(\al, \be ,\ga)^T$, where the superscript $T$ means transpose (thus
$(\al, \be, \ga)$ is a row while $(\al, \be ,\ga)^T$ is a column), and 
\beq M ~=~ \left( \begin{array}{ccc}
2 & -\De & -\De \\
-\De & 2 & -\De \\
-\De & -\De & 2 \end{array} \right). \eeq
The eigenvalues of $M$ are given by $2 - 2 \De$, $2 + \De$, and $2 + \De$.
We see that all the eigenvalues are positive in the range $-2 < \De < 1$
which is therefore the region of stability of phase $II$. For $\De < -2$,
there is a transition to phase $V$, while for $\De > 1$, there is a transition
to phase $I$ as we have already seen.

We will now make a few comments about phase $V$. We find numerically 
that in the classical ground state in this phase, the spins generally vary 
continuously with the parameters $\De$ and $D$; further the spin configuration
on sites $A$, $B$ and $C$ is generally not coplanar (unless $D=0$). Further, 
we have not found an analytical expression for the ground-state spin 
configuration except on the line $D=0$. Guided by numerical results, we find 
that the ground state in the region $\De \le -2$ and $D=0$ is given by $\phi_A 
= \phi_B = \phi_C = 0$, and $\ta_A = \al, ~\ta_B = \pi - \al$ and $\ta_C = 
\pi/2$ (or some permutation of this pattern), where $\al$ lies in the
range $0 \le \al \le \pi/2$ and is given by
\beq \sin \al ~=~ - ~\frac{1}{\De + 1}. \eeq
We see that as $\De$ decreases from $-2$ to $-\infty$, the angles $\ta_i$
vary continuously from $\ta_A = \ta_B = \ta_C = \pi/2$ (which is the same
configuration as in phase $II$) to $\ta_A = 0, ~\ta_B = \pi$ and $\ta_C = 
\pi/2$. Further the energy of a single triangle is given by
\beq E ~=~ JS^2 ~\left( ~\De ~+~ \frac{1}{\De + 1} \right). \label{en2} \eeq
At $\De = -2$, Eq.~\eqref{en2} has the value $-3JS^2$; hence the classical
ground-state energies agree as we approach the line $\De=-2$ from phases $II$
and $V$.

In the following sections we will use spin-wave theory to study the properties 
of magnons in phases $I ~-~ IV$, both in the bulk and at the edges of an 
infinitely long strip of the system.

\section{Spin wave analysis in phase $I$}
\label{sec:phaseI}

The classical ground state in this phase has all the three spins pointing 
along the $+ \hat z$ or $- \hat z$ direction; for definiteness, we will
take them to point along $+ \hat z$. To calculate the magnon spectrum in this 
phase, we use the Holstein-Primakoff transformation~\cite{ande}. 
We write the spin operators in terms of bosonic creation and annihilation 
operators as
\bea A_\vn^z &=& S ~-~ a_\vn^\dg a_\vn, \non \\
A_\vn^+ &=& A_\vn^x + iA_\vn^y ~\simeq~ \sqrt{2S}~a_\vn, \non \\ 
A_\vn^- &=& A_\vn^x + iA_\vn^y ~\simeq~ \sqrt{2S}~a_\vn^\dg, \label{hp} \eea
where $a^\dg_\vn$ and $a_\vn$ are the bosonic creation and annihilation 
operators at the site denoted by $\vn = (n_1,n_2)$. The last two equations 
hold in the limit of large $S$. We have similar transformations for the spin 
operators $B_\vn$ and $C_\vn$. Next, we Fourier transform to momentum space
\beq (a,b,c)_{\vn} ~=~ \displaystyle \sum_{\vk} ~(a,b,c)_{\vk} ~
e^{i\vk\cdot\vn}, \label{FT} \eeq
where the sum over $\vk$ goes over the Brillouin zone (BZ) of the triangular 
lattice formed by the unit cells of the kagome lattice. Retaining terms 
only up to second order in the bosonic operators, 
we obtain a spin-wave Hamiltonian of the form
\beq H ~=~ \displaystyle \sum_{\vk} ~\begin{pmatrix} a^\dg_\vk & b^\dg_\vk &
c^\dg_\vk & \end{pmatrix} ~h(\vk) ~\begin{pmatrix}a_\vk \\ b_\vk \\ c_\vk \\ 
\end{pmatrix}, \label{habc} \eeq
where the $3 \times 3$ matrix $h(\vk)$ has the form
\beq h(\vk) = JS \begin{pmatrix} 4\De &-\mathcal{D} f(-k_1) & -\mathcal{D^*}
f(-k_2) \\ \\ 
-\mathcal{D^*} f(k_1) & 4\De & -\mathcal{D}f(k_1-k_2) \\ \\
-\mathcal{D} f(k_2) & -\mathcal{D^*}f(k_2-k_1) & 4\De \end{pmatrix},
\label{hvk} \eeq
with 
\bea \mathcal{D} &=& 1 ~+~ \frac{i D}{J}, \non \\
f(k) &=& 1+e^{ik} ~=~ f^*(-k). \eea
(We note that there is a factor of $JS$ in Eq.~\eqref{hvk}; hence all the
expressions given below for the magnon energies will be proportional to $JS$).
Note that we have written $h (\vk)$ in terms of the momenta $k_1$ and $k_2$ 
which are along the directions of the unit vectors $\hn_1$ and $\hn_2$
as shown in Fig.~\ref{lattice}. These are related to the $x$ and $y$ momenta 
as $k_x = k_1/\sqrt{3}$ and $k_y =(2 k_2 - k_1)/3$. 

Since the Hamiltonian in Eqs.~(\ref{habc}-\ref{hvk}) is number conserving 
(i.e., there are only terms like $a^\dg a$ and no terms like $a^\dg a^\dg$,
so that the Hamiltonian commutes with $\sum_\vk (a_\vk^\dg a_\vk + b_\vk^\dg 
b_\vk + c_\vk^\dg c_\vk)$), 
the magnon spectrum can be obtained by directly diagonalizing $h(\vk)$.
(This is in contrast to the spin-wave Hamiltonians in phases $II$ and $III$
which have terms like $a^\dg a^\dg$ and are therefore more difficult
to diagonalize as we will see later).
The eigenvalues of $h (\vk)$ are given by solutions to the cubic equation
\beq (\frac{E}{JS} ~- ~4\De)^3 ~+~ a (\frac{E}{JS} ~-~ 4\De) ~+~ b ~=~ 0, 
\label{cubic} \eeq
where
\bea a &=& -2 ~\Big( 1 + \frac{D^2}{J^2} \Big) \Big(3 ~+~ \cos k_1 ~+~ 
\cos k_2 \non \\
&& ~~~~~~~~~~~~~~~~~~~~~~+ ~\cos (k_1-k_2)\Big), \non \\
b &=& 4 \Big( 1 ~+~ \cos k_1 ~+~ \cos k_2 ~+~ \cos (k_1-k_2) \Big). \eea

We can find the boundaries of phase $I$ by examining where the magnon energy 
becomes zero. We find that at $\vk=(0,0)$, $h (\vk)$ in Eq.~\eqref{hvk} has an
eigenvalue equal to $4 JS (\De -1)$ with eigenvector $(1,1,1)^T$, 
and an eigenvalue $JS (4 \De + 2 - 2 \sqrt{3}
|D/J|)$ corresponding to the eigenvector $(1,e^{-i2\pi/3},e^{i2\pi/3})^T$ if 
$D>0$ and $(1,e^{i2\pi/3}, e^{-i2\pi/3})^T$ if $D<0$. We see that these 
eigenvalues touch zero when $\De = 1$ and $|D/J| = (1+2\De)/\sqrt{3}$. 
This gives the boundaries of phase $I$ as shown in Fig.~\ref{phase_diag}.

\subsection{Discrete symmetries}

There is a discrete set of transformations under which the Hamiltonian in 
Eq.~\eqref{hvk} behaves in a simple way. Let us remove the diagonal part given 
by $4 \De \mathbb{1}_3$ (here $\mathbb{1}_3$ is the $3\times3$ identity 
matrix) and denote the rest of the Hamiltonian by $h' (\vk, D/J)$. Namely, 
\beq h'(\vk,\frac{D}{J}) ~=~ h(\vk) ~-~ 4 JS \De ~\mathbb{1}_3. \eeq
We now define a $3 \times 3$ unitary and diagonal matrix $U$ whose diagonal 
entries are given by $(1,e^{i2\pi/3},e^{i4\pi/3})$; this generates a 
finite group with elements given by $(\mathbb{1}_3,U,U^2)$. We then find that
\bea U h' (\vk, \frac{D}{J}) U^{-1} &=& (-\frac{1}{2} + \frac{\sqrt{3}D}{2J})
h'(\vk, \frac{\frac{D}{J} + \sqrt{3}}{1 - \frac{\sqrt{3}D}{J}}), \non \\
U^2 h' (\vk, \frac{D}{J}) U^{-2} &=& (-\frac{1}{2} - \frac{\sqrt{3}D}{2J})
h'(\vk, \frac{\frac{D}{J} - \sqrt{3}}{1 + \frac{\sqrt{3}D}{J}}). \non \\
&& \label{hz3} \eea
[Since 
\beq \tan^{-1} \left( \frac{D/J \pm \sqrt{3}}{1 \mp \sqrt{3} D/J} \right) ~=~
\tan^{-1} (D/J) ~\pm~ \frac{\pi}{3}, \eeq
we can interpret the transformations in Eq.~\eqref{hz3} as rotations by 
$\pm \pi/3$
in the two-dimensional plane defined by points with coordinates $(1,D/J)$].
Since the eigenvalues of a matrix are invariant under unitary transformations,
Eq.~\eqref{hz3} implies that the eigenvalues of $h'(\vk,D/J)$, denoted as
$E'_i (\vk,D/J)$ (where $i=1,2,3$ denotes the three bands), satisfy the 
relations
\bea E'_i (\vk, \frac{D}{J}) &=& (-\frac{1}{2} + \frac{\sqrt{3}D}{2J})~
E'_i ( \vk, \frac{\frac{D}{J} + \sqrt{3}}{1 - \frac{\sqrt{3}D}{J}} ) \non \\
&=& (-\frac{1}{2} - \frac{\sqrt{3}D}{2J}) ~E'_i ( \vk, \frac{\frac{D}{J} -
\sqrt{3}}{1 + \frac{\sqrt{3}D}{J}} ). \label{ez3} \eea
Thus there are sets of three values of $D/J$ such that the energies $E'_i (\vk,
D/J)$ are related to each other by some constants which do not depend on $\vk$;
such sets of values of $D/J$ will therefore share some properties of the energy
bands (for instance, two of the bands touching each other at some value of 
$\vk$). In particular, setting $D/J=0$, we see that
\bea E'_i (\vk, 0) &=& -\frac{1}{2} ~E'_i (\vk, \sqrt{3}) \non \\
&=& -\frac{1}{2} ~E'_i (\vk, - \sqrt{3}). \label{ez32} \eea
We will see below that there is no gap between the bottom and middle bands 
and also between the middle and top bands at both $D/J=0$ and $\pm \sqrt{3}$; 
this is consistent with Eq.~\eqref{ez32}.

\subsection{Energy spectrum of magnons}

In phase $I$, the symmetry of the Hamiltonian under spin rotations in the
$x-y$ plane is not broken by the ground state. We therefore do not expect
a zero energy Goldstone mode. We indeed find that the magnon states have
strictly positive energy for all momenta $\vk$. 

In the absence of the DMI, i.e., for $D=0$, we get a dispersionless top band. 
Further, we find that there are no gaps between any pair of bands. The bottom 
and middle bands (labeled 1 and 2 respectively) touch at two momenta in the 
Brillouin zone, $\vk = \pm (2\pi/(3\sqrt{3}), -2\pi/3)$, while the middle and 
top (labeled 3) bands touch at $\vk = (0,0)$. The spectrum is also gapless at 
$D/J = \pm \sqrt{3}$. However, the gap closing points are now inverted, 
i.e., bands 1 and 2 touch at $\vk = (0,0)$, whereas bands 2 and 3 touch at 
$\vk = \pm (2\pi/(3\sqrt{3}), -2\pi/3)$; further, the bottom band is 
dispersionless in this case. 

Figures~\ref{spec_ph1a} and \ref{spec_ph1b} show the bulk energy dispersion 
for $\De = 2$ and $D/J = 2/\sqrt{3}$ and $4/\sqrt{3}$ respectively. The 
reason for choosing these parameter values will be made clear in the following 
discussion about the variation of the energy gaps with the DMI strength.
{Figures~\ref{edge_ph1a} and \ref{edge_ph1b} also show the spectrum of 
edge states. [By edge states, we mean states which are localized near either 
the top edge or the bottom edge; see Fig.~\ref{lattice}. The wave functions of 
these states are plane waves along the $\hat x$ direction since the system has 
translation symmetry in that direction, but they decay exponentially as we 
move away from the edges along the $\hat y$ direction.]}
We see in Figs.~\ref{edge_ph1a} and \ref{edge_ph1b} that there is no 
$k_x \to - k_x$ symmetry in the spectrum of edge states; this is because the 
DMI in the Hamiltonian in Eq.~\eqref{Hamiltonian} does not have a $x \to - x$ 
symmetry. On the other hand, the Hamiltonian is invariant if we change both 
$x \to - x$ and $y \to - y$ simultaneously; hence the bulk spectrum is 
invariant under a simultaneous change of $k_x \to - k_x$ and $k_y \to - k_y$. 
Since the bulk bands shown in Figs.~\ref{edge_ph1a} and \ref{edge_ph1b} 
combine all values of $k_y$ for each value of $k_x$, they show a $k_x \to - 
k_x$ symmetry.

\begin{widetext}
\begin{center}
\begin{figure}[H]
\begin{center}
\subfigure[]{\includegraphics[width=7cm]{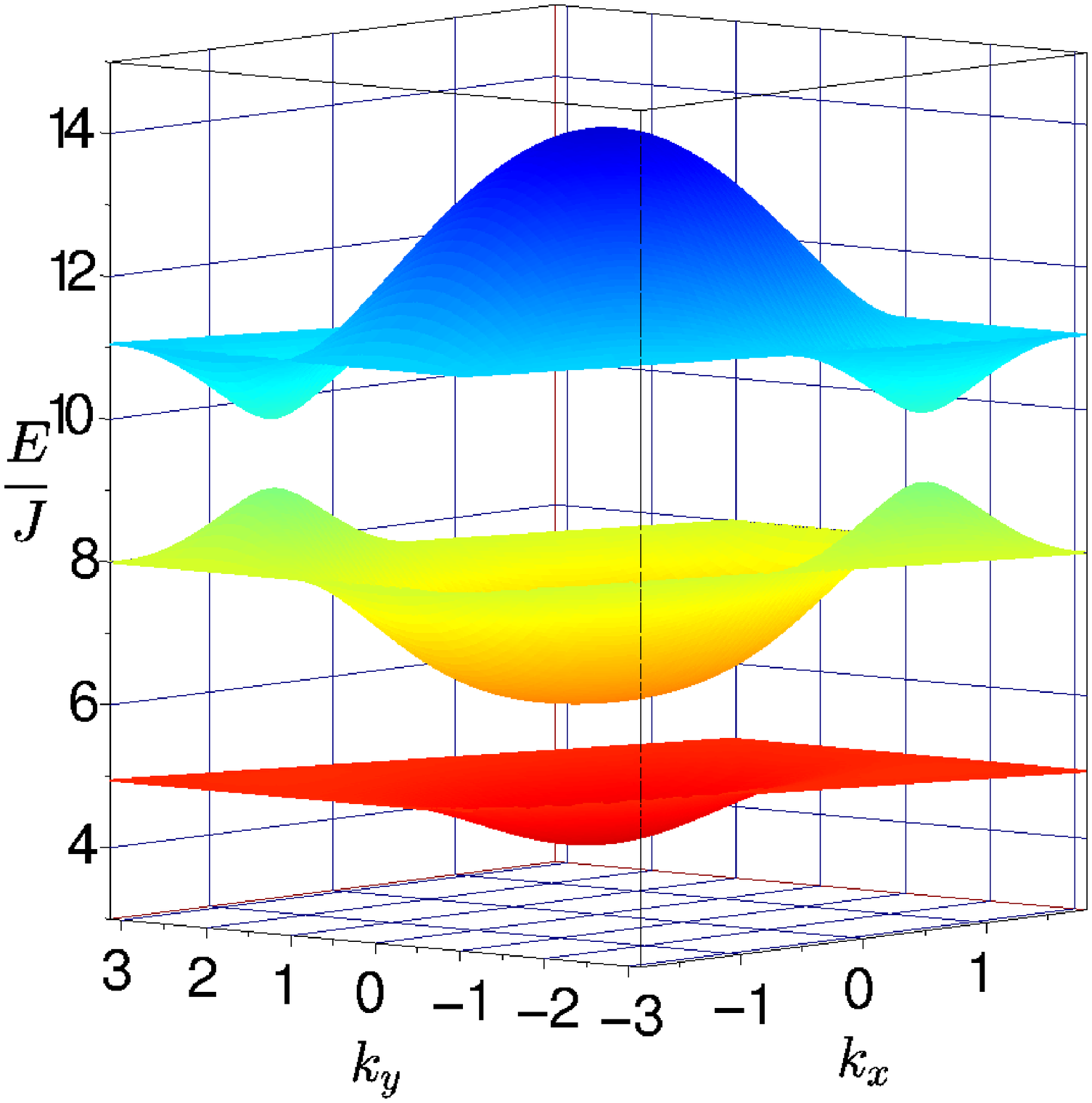}
\label{spec_ph1a}}
\subfigure[]{\includegraphics[width=7cm]{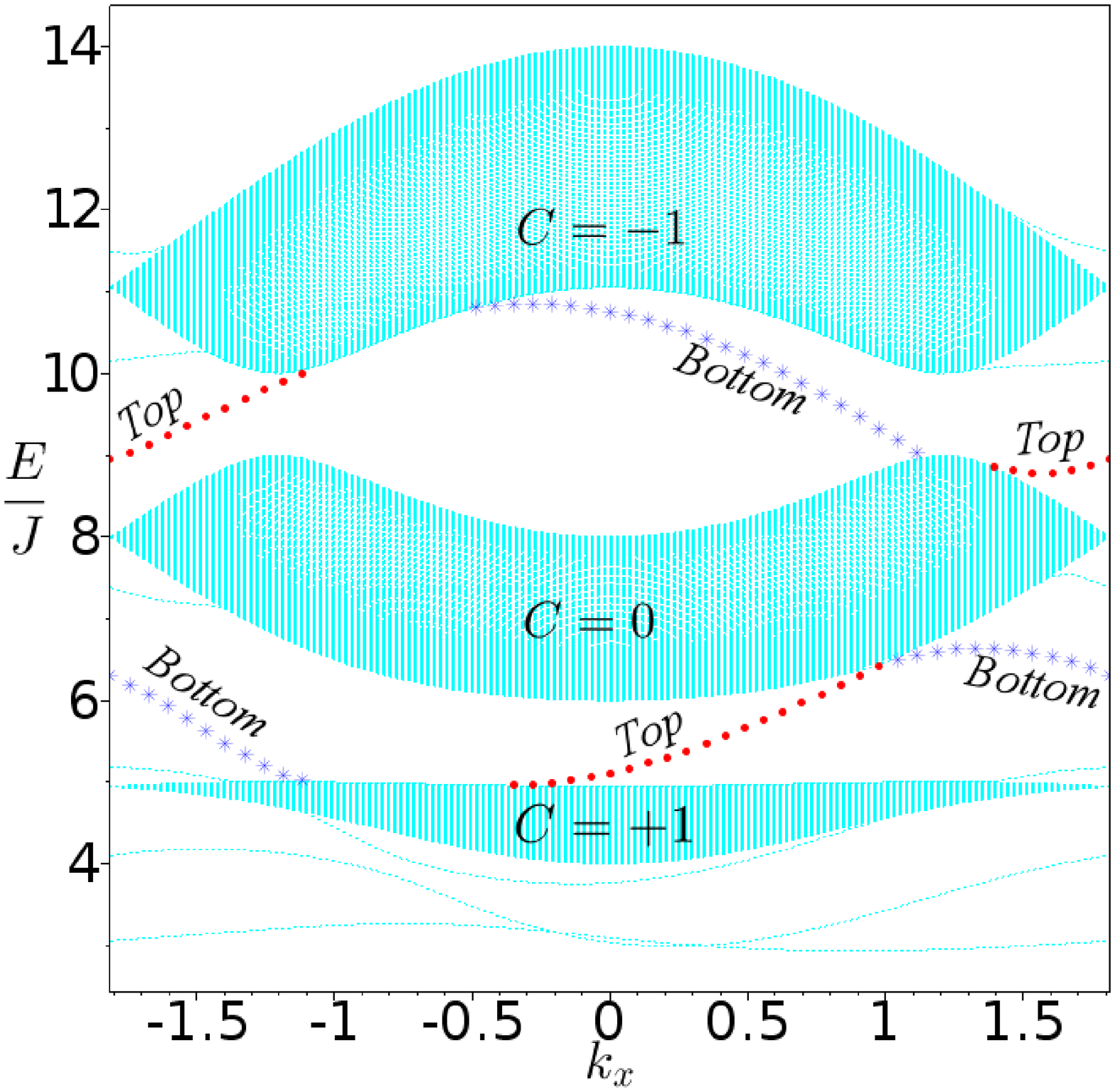}
\label{edge_ph1a}} \\
\subfigure[]{\includegraphics[width=7cm]{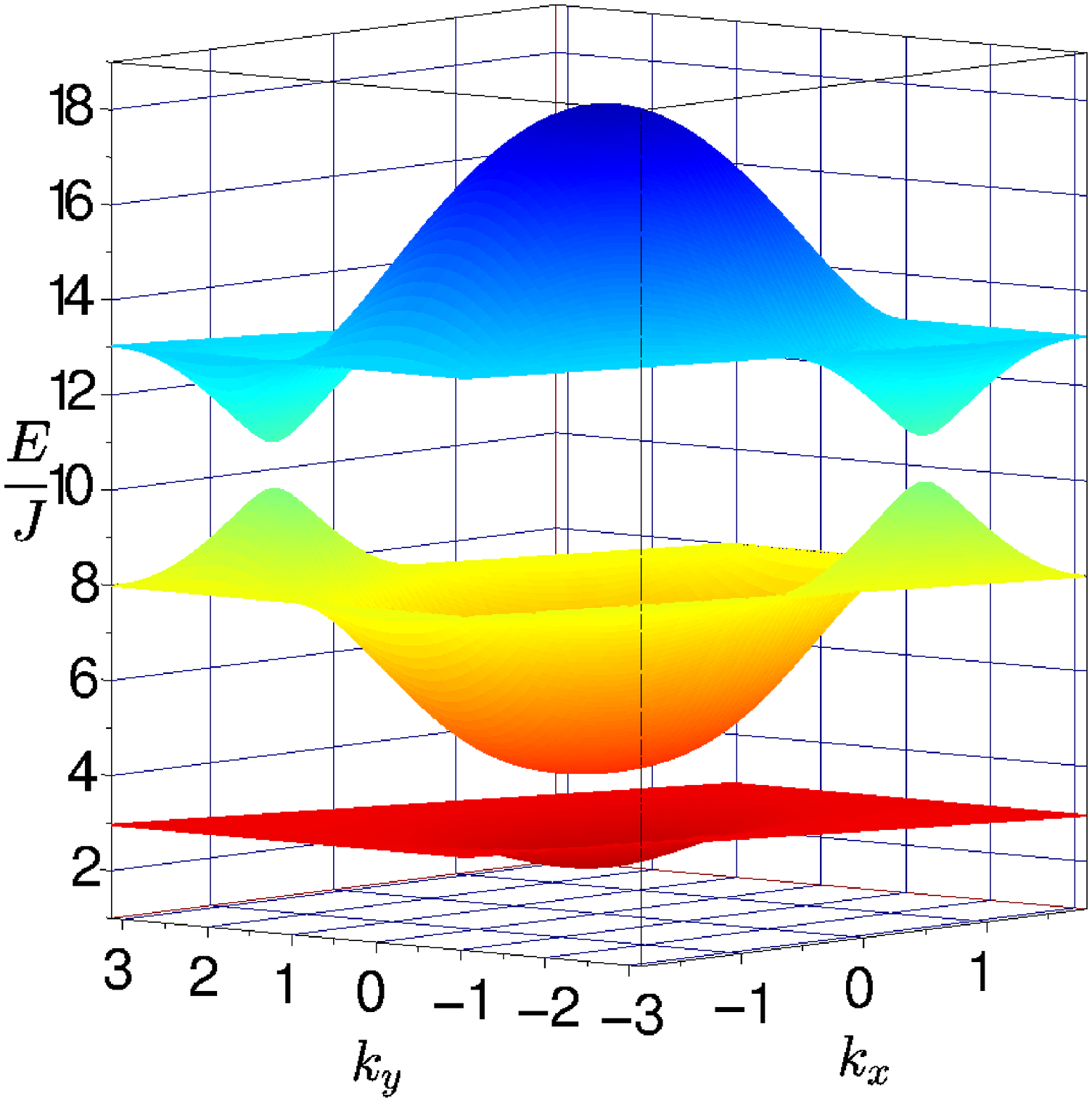}
\label{spec_ph1b}}
\subfigure[]{\includegraphics[width=7cm]{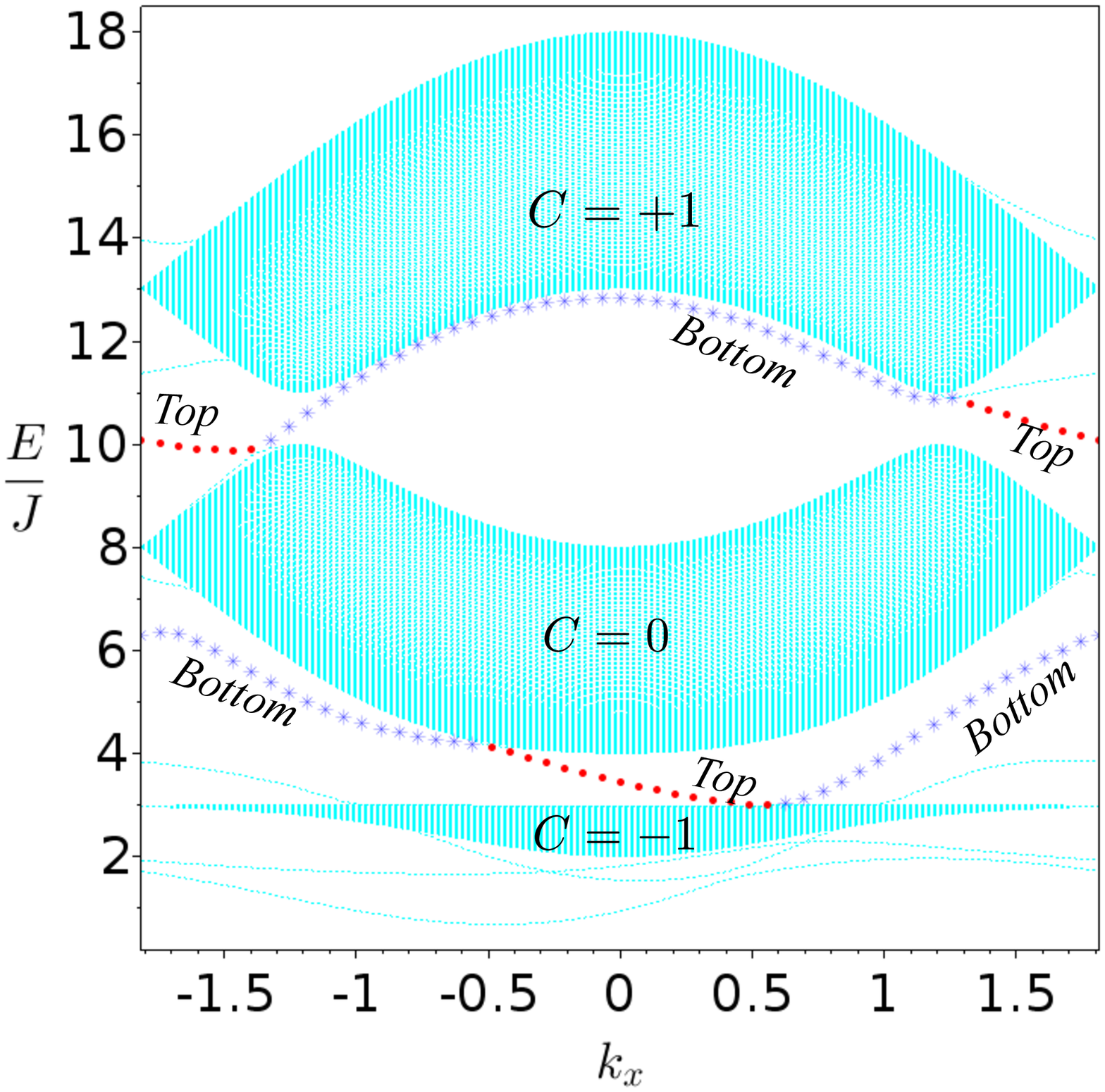}
\label{edge_ph1b}}
\caption{Magnon energy dispersions in the bulk (a,c) and at the edges (b,d) 
for an infinitely long strip as shown in Fig.~\ref{lattice}, in phase $I$ with 
$\De = 2$. We have taken $D/J = 2/\sqrt{3}$ and $4/\sqrt{3}$ in the top and 
bottom rows respectively. For $D/J = 2/\sqrt{3}$ the Chern numbers for the 
bands with increasing energy are $(1,0,-1)$, while for $D/J = 4/\sqrt{3}$ the 
Chern numbers are $(-1,0,1)$. A phase transition occurs at $D/J = 
\sqrt{3}$ where the gaps between the bands close (Fig.~\ref{gap_vs_DM}) and 
the Chern number becomes ill-defined. The red (dotted) lines, marked as ``Top",
show edge states localized near the top edge whereas the blue (asterisks) 
lines, marked as ``Bottom", show edge states at the bottom edge. There are 
also some additional modes in the gap which are drawn in light blue. However 
these states do not connect one band to the other and are not topologically 
protected. Thus any small perturbation in the Hamiltonian can modify them 
significantly and even merge them with the bulk bands that they are closest 
to.}
\end{center}
\end{figure}
\end{center}
\end{widetext}

\begin{figure}
\includegraphics[height=6cm]{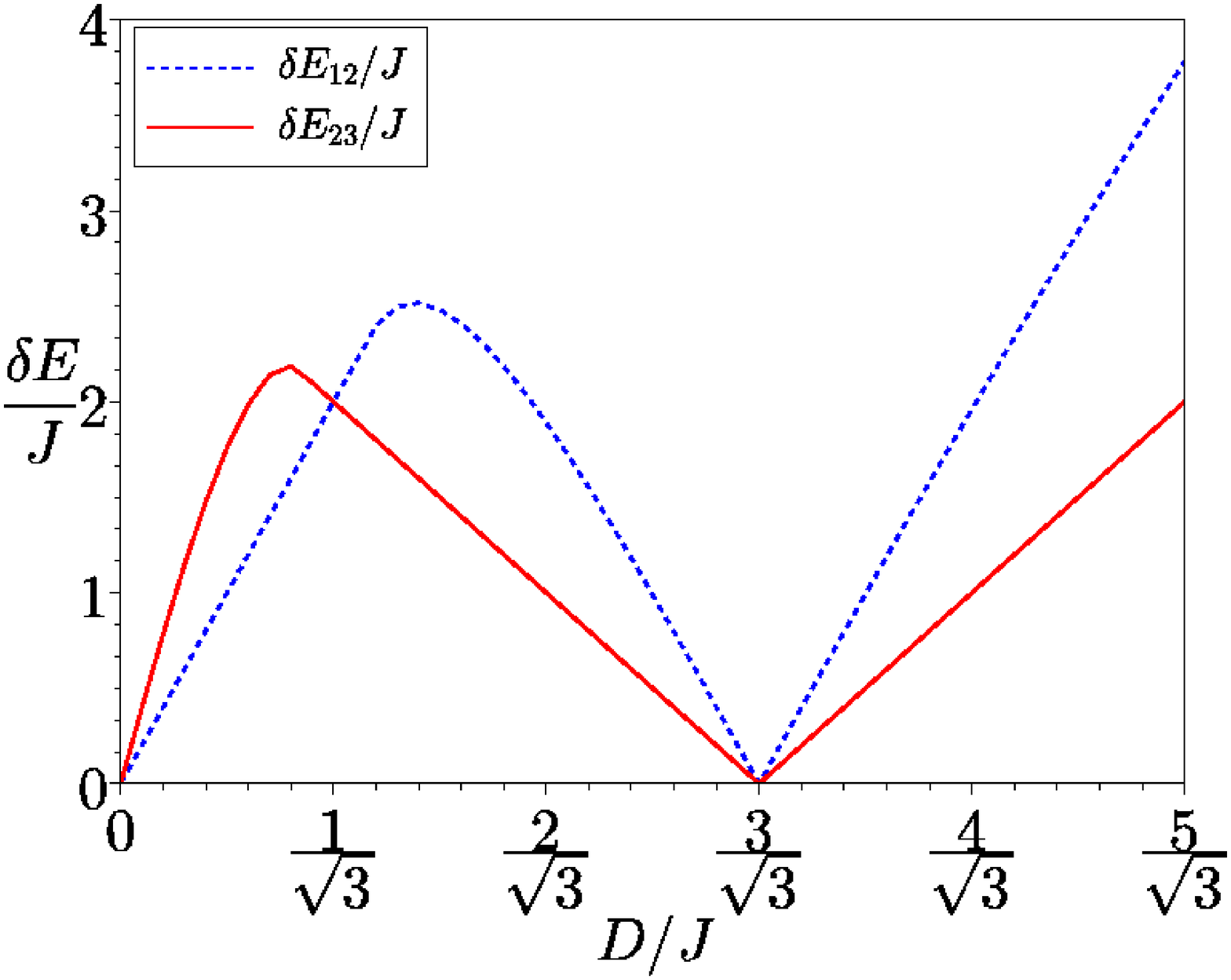}
\caption{Gap between bottom and middle bands, $\de E_{12}/J$, and between
middle and top bands, $\de E_{23}/J$, as a function of $D/J$ in the 
ferromagnetic phase in which all the spins point along the $+ \hat z$ or
$-\hat z$ direction. The gaps vanish at $D/J = 0$ and $\sqrt{3}$. We have 
taken $\De = 2$.} \label{gap_vs_DM} \end{figure}

We plot the DMI dependence of the energy gaps $\de E_{12}$ (between the 
bottom and middle bands) and $\de E_{23}$ (between the middle and top bands)
for $\De =2$ in Fig.~\ref{gap_vs_DM}. We find that as we increase the DMI 
strength $D$ from zero both the gaps initially increase and then decrease. 
At $D/J = \sqrt{3}$, the gaps close, at the same momenta as in the case of 
$D = 0$, before the gaps open again. For negative $D$, the plot is given by a 
mirror reflection about the line $D=0$, i.e., the gaps depend only on the 
magnitude of $D$ and not on its sign. Thus the gaps close at $D/J = -\sqrt{3}$ 
too. Note that the gaps are completely independent of the anisotropy ratio 
$\De$ since this appears only in the diagonal elements of $h(\vk)$ and 
therefore only shifts all the energy levels by $4\De$. 

Since there are gaps between all pairs of bands if $D/J \ne 0, ~\pm \sqrt{3}$,
we can calculate topological 
invariants for the system, namely, the Chern numbers of all the bands. 
These Chern numbers $C_i$'s ($i=1,2,3$ for the bottom, middle and top bands 
respectively) are shown in Figs.~\ref{edge_ph1a} and \ref{edge_ph1b} for 
$D/J = 2/\sqrt{3}$ and $4/\sqrt{3}$. In order to calculate $C_i$'s, we first 
define the Berry curvature in band $i$ using the normalized eigenstates 
$\psi_i(\vk)$'s and eigenvalues $E_i(\vk)$'s of the Hamiltonian:
\beq \Omega_i (\vk) = {\text i} \displaystyle \sum_{j \neq i} 
\frac{(\psi^\dag_i(\vk) {\vec \nabla}_\vk H(\vk)\psi_j(\vk))\times
(\psi^\dag_j(\vk) {\vec \nabla}_\vk H(\vk)\psi_i(\vk))}{[E_i(\vk) - 
E_j(\vk)]^2}. \eeq
The Chern number for band $i$ can then be calculated by integrating the Berry 
curvature over the Brillouin zone, i.e.,
\beq C_i ~=~ \frac{1}{2\pi} \int_{BZ} d^2k ~ ~\Omega_i (\vk). \eeq
In this paper we have used the method in Ref.~\onlinecite{fuku} to calculate 
these topological invariants.

The topological character of the bands changes whenever the band gaps close 
and open. For $0<D/J<\sqrt{3}$ the bottom middle and top bands have Chern 
numbers equal to $(1,0,-1)$, whereas for $D/J >\sqrt{3}$ these numbers 
change to $(-1,0,1)$ and remain so till we reach the phase transition line with
phase $III$. Since the band gap closes at $D/J = \sqrt{3}$ the Chern numbers 
are ill-defined there. A similar behavior is seen for $D<0$. The Chern numbers 
are $(-1,0,1)$ for $-\sqrt{3} < D/J < 0$ and change to $(1,0,-1)$ when 
$D/J < -\sqrt{3}$ till we reach the transition to phase $IV$. We will 
now study if the Chern numbers have any implications for the 
edge states of the system. 

\red{We can understand the existence of these topological phases through the 
following symmetry arguments. In the absence of the DMI term (i.e., for $D=0$),
the Hamiltonian in Eq.~\eqref{hvk} satisfies $h(\vk) = h^*(-\vk)$. 
This is a pseudo-time-reversal symmetry similar to the one discussed in 
Ref.~\onlinecite{shind}. Switching on a finite value of $D$ breaks this 
pseudo-time-reversal symmetry as well as the reflection symmetry mentioned 
after Eq.~\eqref{Hamiltonian}; this gives a chirality or handedness to the 
system and thereby produces a topological phase.} 

\red{It is interesting to note that for a finite value of $D$,
the Hamiltonian in Eq.~\eqref{hvk} satisfies $h(-\vk,-D/J)=h^* (\vk,D/J)$.
The eigenstate of $h(\vk,D/J)$ in band $i$ therefore satisfies $\psi_i (-\vk,
-D/J) = \psi_i^* (\vk,D/J)$. From this we can show that the Berry curvature
and Chern number satisfy $\Omega_i (-\vk,-D/J) = -\Omega_i (\vk,D/J)$ and 
$C_i (-D/J) = - C_i (D/J)$.}

We consider a strip of the kagome lattice which is infinitely long in the 
$\hat x$ direction (this is the same as the $\hat{n}_1$ direction). This makes 
$k_x$ (and hence $k_1$) a good quantum number. We take the width of the strip 
$N_2$ in the $\hat{n}_2$ direction to be about $150$ unit cells; each unit 
cell consists of a triangle of three spins as usual; see Fig.~\ref{lattice}. 
This reduces the lattice to a series of one-dimensional chains; each chain
has $N_2$ unit cells (and therefore $3N_2$ sites) and is coupled to its 
neighboring chains. Each chain is labeled by the value of $n_1$, and the 
magnon wave functions at all the sites of the chain carry a plane wave factor 
equal to $e^{ik_1 n_1}$. The wave functions on adjacent chains are therefore 
related to each other by plane wave factors equal to $e^{\pm ik_1}$. We can 
then construct a $3N_2 \times 3N_2$ Hamiltonian for each value of $k_1$ 
whose eigenvalues are the energies at that $k_1$. The spectra of edge 
states are shown in Figs.~\ref{edge_ph1a} and \ref{edge_ph1b} for $\De = 2$
and $D/J = 2/\sqrt{3}$ and $4/\sqrt{3}$ respectively. Note that the energy 
spectra are quite different for states localized near the top and bottom 
edges; this is because the shapes of these edges are different from each 
other as we see in Fig.~\ref{lattice}. Indeed, one of our motivations
for choosing the top and bottom edges to have different shapes was precisely
to test if the number of states on the two edges are identical if the system 
is in a topological phase. It is the hallmark of a system in a topological
phase that the number of states is the same on any edge regardless of
its shape, assuming that the edge are sufficiently far from each other so that
states on different edges do not hybridize with each other.

Since the system is topological away from the lines $D/J =0, ~\pm \sqrt{3}$, 
the edge states are topologically protected. We find that the number of edge 
states $\nu_i$ at either the top edge or the bottom edge with energies in the 
$i$-th band gap is related to the Chern number up to the $i$-th band 
as~\cite{hats1,hats2,mook2}
\beq \nu_i ~=~ \Bigg{|} \sum_{j\leq i}C_j \Bigg{|}. \label{nui} \eeq
This is reflected in the edge state spectra in Figs.~\ref{edge_ph1a} and 
\ref{edge_ph1b}. We note that Eq.~\eqref{nui} only involves the number 
of edge states, not their momentum. This is not unusual; edge states need not 
exist for the full range of values of $k_x$ which are allowed for the bulk 
states. (We can see this in a completely different system which is also in 
a topological phase; see Ref.~\onlinecite{kane}). Equation~\eqref{nui} implies
that between the lower and and middle bands, there should be one branch of 
edge state at both the top and bottom edges, and the same should be true 
between the middle and top bands. This is indeed what we observe in
Figs.~\ref{edge_ph1a} and \ref{edge_ph1b}.

To summarize, the system in phase $I$ is not topological in the absence 
of DMI, i.e., if $D = 0$ (this is clear since we can have a topological 
system only if there are terms which break the symmetry under reflection 
about the $x$ or $y$ axis). Then the symmetry of the magnon spectrum 
under $\pm \pi/3$ phase rotations of the complex variable ${\cal D} = 1 + 
iD/J$ (as discussed in Sec. III A) shows that the system must also be 
nontopological when $D/J = \pm \sqrt{3}$. These three values of $D/J$ (namely, 
zero and $\pm \sqrt{3}$) give the transition lines between the four 
topological phases.

\subsection{Thermal Hall Conductivity}

The thermal Hall conductivity $\ka^{xy}$ of the system is closely related to
Berry curvature $\Omega_i (\vk)$ of the $i$-th band of the system 
as~\cite{mats1,mats2}
\bea \ka^{xy} ~=~ -~\frac{k_B^2 T}{4\pi^2 \hbar} ~\displaystyle\sum_{i} 
\displaystyle \int_{BZ} d^2k ~c_2(\rho_i (\vk)) ~\Omega_i (\vk), \label{kxy} 
\eea
where $T$ is the temperature, $k_B$ is the Boltzmann constant, $\hbar$ is 
Planck's constant, and $i$ denotes the band index, i.e., $i=1,2$ and $3$ for 
the bottom, middle and top bands respectively. In Eq.~\eqref{kxy}, the 
function $c_2$ is given by 
\bea c_2(x) ~=~ (1+x){\Big(}\ln\frac{1+x}{x}{\Big)}^2 ~-~ (\ln x)^2 ~-~ 
2{\text{Li}}_2(-x), \non \\ \eea
where ${\text{Li}}_2$ is the dilogarithm function
\bea {\text{Li}}_2(z) ~=~ - ~\int_0^z du ~\frac{\ln(1-u)}{u}, \eea
and the argument of $c_2$ is $\rho_i (\vk)$ which is equal to
the Bose distribution function for the energy $E_i (\vk)$, i.e.,
\bea \rho_i (\vk) ~=~ \frac{1}{e^{E_i (\vk)/(k_B T)} ~-~ 1}. \eea
Since the thermal Hall conductivity directly depends on the Berry curvature, 
we expect it to behave differently in different topological phases.

\begin{figure}[h!]
\subfigure[]{\includegraphics[width=8cm]{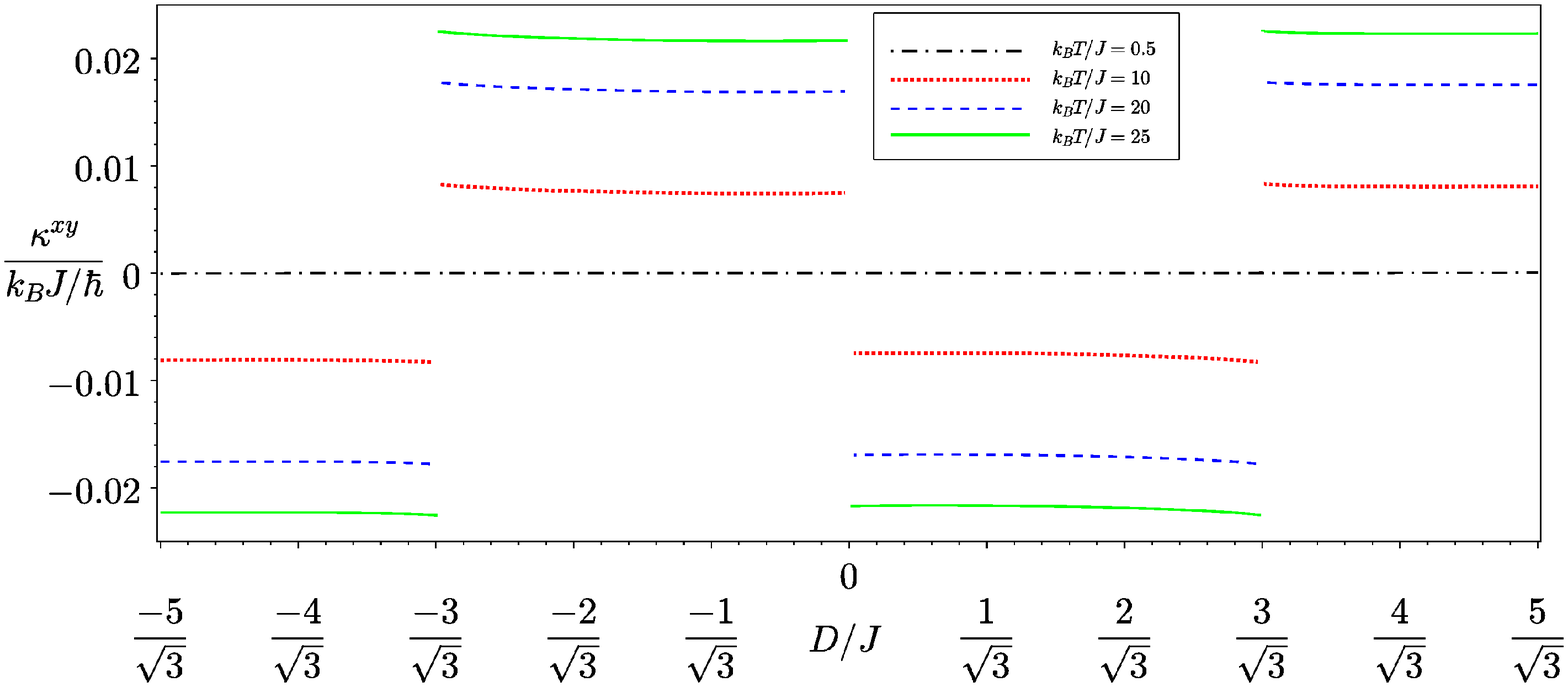}\label{sfig.hallD}}\\
\subfigure[]{\includegraphics[width=8cm]{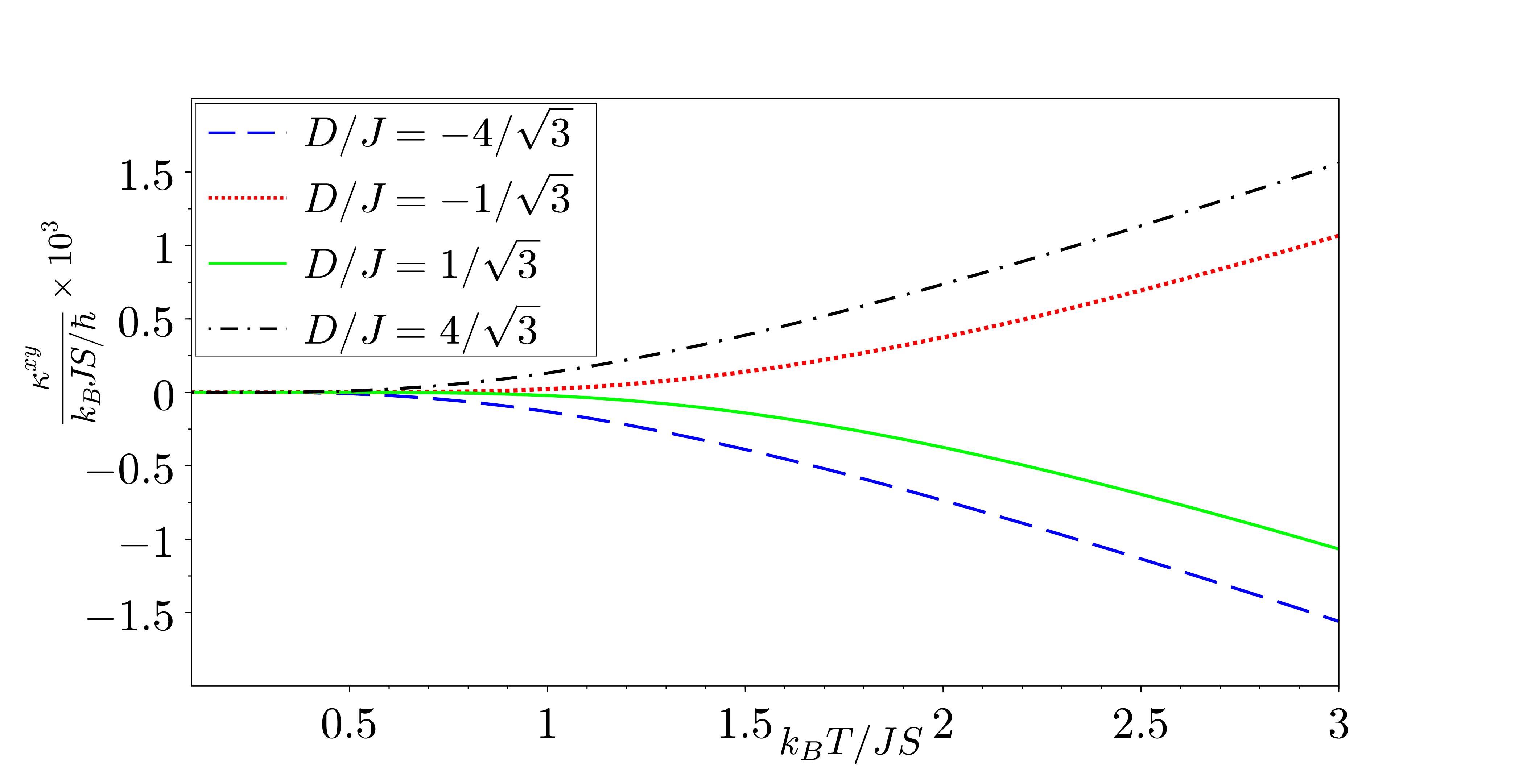}\label{sfig.hallT}}\\
\subfigure[]{\includegraphics[width=9cm]{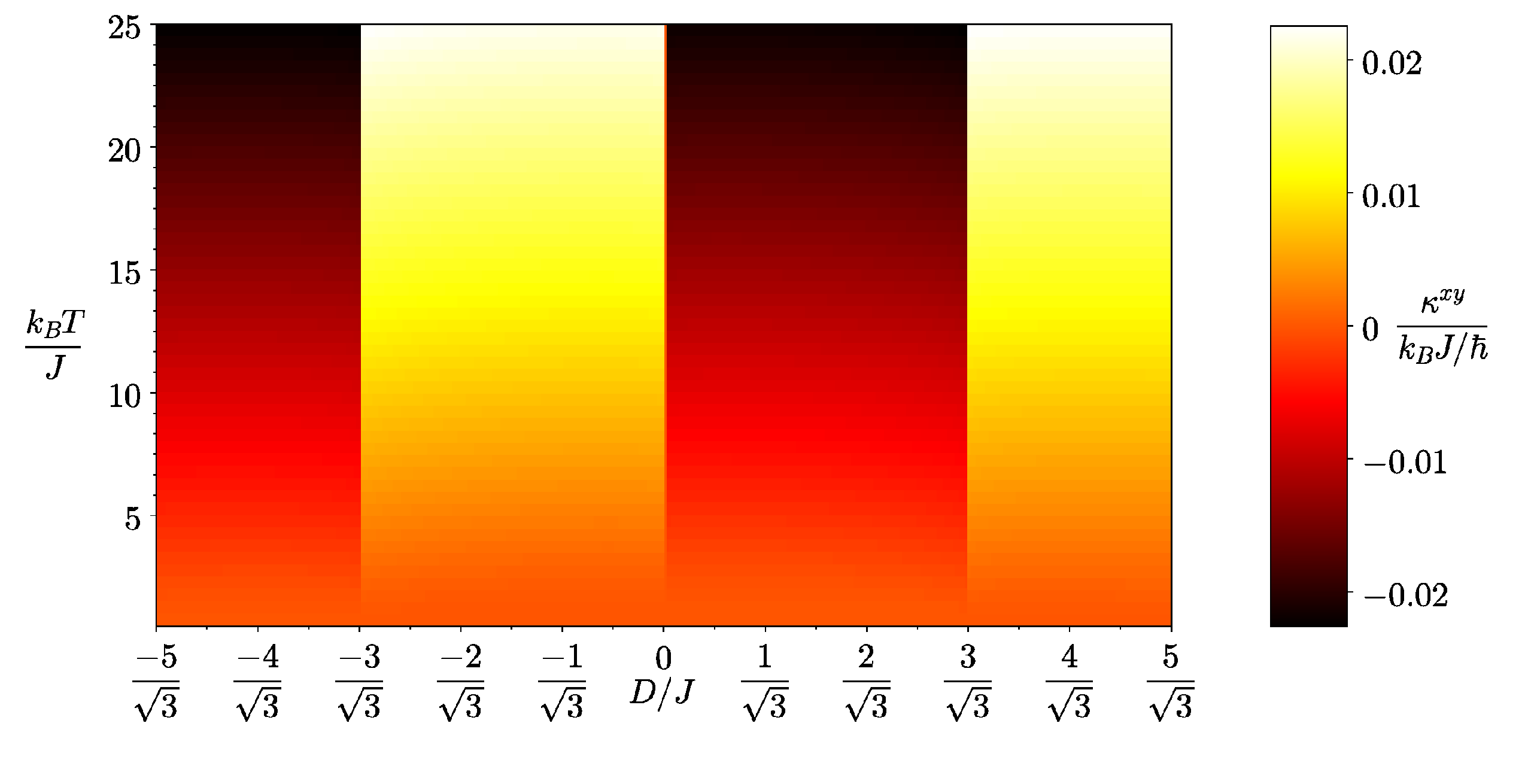}\label{sfig.hallsurf}}
\caption{Variation of $\ka^{xy}$ with temperature $T$ and DMI strength $D$, 
for $\De = 2$. (a) Varying $D/J$ while keeping $T$ constant, we find
that at every phase transition point, i.e., at $D/J = 0,\pm\sqrt{3}$, the
Hall conductivity $\ka^{xy}$ changes sign. (b) For a given value of $D/J$,
$\ka^{xy}$ increases monotonically with temperature. We have used four
different values of $D/J$ as shown in the legend. (c) Color plot of
$\ka^{xy}$ vs $D/J$ and $k_BT/JS$. The darker (red) regions where
$\ka^{xy}$ becomes more and more negative with increasing temperature are
where the Chern numbers are $(+1,0,-1)$, whereas the lighter (yellow) regions
where $\ka^{xy}$ increases with temperature are where the Chern numbers are
$(-1,0,+1)$.} \label{fig:hall} \end{figure}

Figure~\ref{fig:hall} shows the variation of $\ka^{xy}$ with the temperature
$T$ and the DMI strength $D$. As shown in Fig.~\ref{sfig.hallD}, whenever 
we cross a topological phase boundary, i.e., at $D/J = 0, \pm\sqrt{3}$,
there is a discontinuity in $\ka^{xy}$. These are exactly the points where 
the Chern numbers of the bands change. Since $\ka^{xy}$ flips sign at these 
points, the discontinuity is larger at higher temperatures.

We note that $\ka^{xy}$ is a monotonic function of $T$ for a given value
of $D$. For $D/J = -4/\sqrt(3)$ and $1/\sqrt{3}$, the $\ka^{xy}$
decreases and goes as $-T\log(T)$ for large $T$. At both these values
of $D/J$ the three energy bands have Chern numbers $(+1,0,-1)$. Similarly, 
when $D/J = -1/\sqrt{3}$ or $4/\sqrt{3}$, the bands have the opposite Chern
numbers and $\ka^{xy}$ now increases as $T\log(T)$ with increasing $T$. 
This behavior is shown in Fig.~\ref{sfig.hallT}. For all values of $D$,
$\ka^{xy} \to 0$ exponentially fast as $T \to 0$.

The surface plot Fig.~\ref{sfig.hallsurf} makes the connection between Chern
number and Hall conductivity clearer. The regions where the Chern numbers are
$(+1,0,-1)$, i.e., for $-5/\sqrt{3}<D/J<-\sqrt{3}$ or $0<D/J<\sqrt{3}$,
$\ka^{xy}$ decreases with increasing temperature $T$. These are the darker
(red) regions in the figure. On the other hand, when $-\sqrt{3}<D/J<0$ or 
$\sqrt{3}<D/J<5/\sqrt{3}$ and the Chern numbers are $(-1,0,+1)$, $\ka^{xy}$ 
increases with $T$. The lighter (yellow) regions in the figure show the 
behavior of $\ka^{xy}$ in this regime.

In Fig.~\ref{fig:hall}, we have only shown results up to a temperature 
given by $k_B T = 3 JS$. This is because the magnons have a maximum band width 
of about $3JS$; for temperatures larger than this, a large number of magnons 
would become thermally excited and we would have to take their interactions 
into account, which would require us to go to cubic and higher orders in the 
$1/S$ expansion of the Hamiltonian.

We can obtain an estimate of $\ka^{xy}$ assuming some typical values 
of various parameters. Taking $J \sim 1$ 
meV, $T \sim 10$ K and $S = 1/2$, we see from the red dashed line ($k_B T/JS 
= 2$) in Fig.~\ref{sfig.hallD} that $\ka^{xy}$ is of the order of $10^{-13}$
W/K for a range of values of $\De$. However it is clear from that figure that 
$\ka^{xy}$ varies significantly as the ratio $T/JS$ is changed.

\subsection{Phase $I$ for finite spin}

An interesting fact about phase $I$, which is not true for phases $II ~-~ V$,
is that all the results presented above remain valid for {\it any} value 
of the spin $S$, not just large $S$. Namely, both the ground state and
the one-magnon states that we have discussed above are exact eigenstates
of the Hamiltonian $H$ in Eq.~\eqref{Hamiltonian} for any value of $S$,
including spin-1/2. To see this we first note that the 
total spin component along the $\hat z$ direction, namely, 
\beq S^z ~=~ \sum_\vn ~(A_\vn^z ~+~ B_\vn^z ~+~ C_\vn^z) \eeq
commutes with $H$ and can 
therefore be used to label the eigenstates of $H$. Next, the sector with the 
maximum or minimum possible value of $S^z$, namely, the state in which all the 
spins have $S_i^z = S \hbar$ (or $- S \hbar$) is an exact eigenstate of $H$; 
here $i$ includes both the unit cell index $\vn$ and the site index $A,B,C$. 
Next, we can find the region where this state is the ground state of $H$ as 
follows. We calculate the energy of the states in the sector in which one of 
the spins has $S_i^z = (S-1) \hbar$ and all the other spins have $S_i^z = S 
\hbar$; we will call this the one-magnon sector. In this sector, the magnon 
annihilation and creation operators $a_\vn$ and $a_\vn^\dg$ are exactly given 
by $A_\vn^+ = \sqrt{2S} a_\vn$ and $A_\vn^- = \sqrt{2S} a_\vn^\dg$, and
similarly, for $b_\vn, ~c_\vn$, etc. We then find that the Hamiltonian for a 
magnon is given by Eqs.~(\ref{habc}-\ref{hvk}) exactly, i.e., with no
corrections at higher orders in a $1/S$ expansion. The form of the magnon 
spectrum and the locations of the boundaries of phase $I$ are therefore 
identical to what we found in the large $S$ limit. The existence of four 
distinct topological phases within phase $I$ also holds for all values of $S$.

\section{Spin wave analysis in phases ~~~~~~~~~~~~~~~~~~~ $II$ and $III$}
\label{sec:phaseII}

In both phases $II$ and $III$, all the spins lie in the same plane in the 
classical ground state. In phase $II$ ($-2 \leq \De \leq 1$, $|D/J| \leq 
\sqrt{3}$), the classical ground state has all the spins pointing along the 
$\hat x$ direction. For the spin-wave analysis, we use the following 
Holstein-Primakoff transformation:
\bea A^x = S-a^\dg a, ~~ A^+ \simeq \sqrt{2S}~a, ~~ A^- \simeq 
\sqrt{2S}~a^\dg, \non \\
B^x = S-b^\dg b, ~~ B^+ \simeq \sqrt{2S}~b, ~~ B^- \simeq \sqrt{2S}~b^\dg, 
\non \\
C^x = S-c^\dg c, ~~ C^+ \simeq \sqrt{2S}~c, ~~ C^- \simeq \sqrt{2S}~c^\dg, 
\label{HP23}\eea
where 
\beq (A,B,C)^{\pm} ~=~ (A,B,C)^y ~\pm~ i ~(A,B,C)^z \label{xx2} \eeq
at every site labeled by $\vec{n} = (n_1,n_2)$.\\

In the ground state of phase $III$, the spins at sites $A$, 
$B$ and $C$ are rotated at an angle $2\pi/3$ with respect to one another as 
shown in Fig.~\ref{classical}. The Holstein-Primakoff transformation in
this case is similar to Eqs.~(\ref{HP23}-\ref{xx2}), except that
\bea A^\pm &=& A^{-\hn_2} \pm i ~A^{-\hn_2'}, \non \\
B^\pm &=& B^{\hn_3} \pm i ~B^{\hn_3'}, \non \\
C^\pm &=& C^{y} \pm i ~C^{z}, \label{HP3} \eea
where $\hn_1$ and $\hn_2$ are the directions along the $A-C$ and $A-B$ bonds of
each unit cell of the lattice (as shown in Fig.~\ref{classical}), $\hn_3$ 
points along $\hn_2-\hn_1$, and the unit vector $\hn'_i$ (where $i=2,3$) is 
orthogonal to $\hn_i$ such that $\hn'_i \times \hn_i$ point along $\hat z$. 
We have not transformed the spin at site $C$ into the new basis as this spin 
points along the $+\hat x$ direction just as in phase $II$. The calculation 
is explained in more detail in Appendix \ref{app.HPIII}.

Using these transformations in the Hamiltonian in Eq.~\eqref{Hamiltonian} and 
substituting the Fourier transforms of the bosonic operators from 
Eq.~\eqref{FT} we get the following general form of the Hamiltonian 
common to both phases $II$ and $III$, 
\beq H(\vk) ~=~ \displaystyle \sum_{\vk} ~\Psi^\dg (\vk) ~\clyM(\vk)~
\Psi^\dg (\vk), \label{ham23} \eeq
where $\Psi$ is a six-component column given by
\beq \Psi^\dg (\vk) ~=~ \begin{pmatrix}a^\dg_\vk & b^\dg_\vk & c^\dg_\vk & 
a_{-\vk} & b_{-\vk} & c_{-\vk} \end{pmatrix}, \eeq
and $M(\vk)$ is a $6 \times 6$ Hermitian matrix of the form 
\beq \clyM(\vk) = JS \Big[ \Big( 4\ga \mathbb{1}_{3} - \al F(\vk) \Big) \otimes
\mathbb{1}_{2} ~-~ \be F(\vk) \otimes \si_x \Big], \eeq
where $\mathbb{1}_n$ denotes the $n\times n$ identity matrix, the Pauli 
matrices are
\beq \si_x = \begin{pmatrix}0 & 1 \\ 1 & 0\end{pmatrix}, ~~\si_y = 
\begin{pmatrix}0 & -i \\ 
i & 0\end{pmatrix}, ~~\si_z = \begin{pmatrix}1 & 0 \\ 0 & -1 \end{pmatrix}, 
\\ \eeq
and
\beq F(\vk) ~=~ \begin{pmatrix} 0 & f(-k_1) & f(-k_2) \\ \\ 
f(k_1) & 0 & f(k_1-k_2) \\ \\
f(k_2) & f(k_2-k_1) & 0
\end{pmatrix}. \label{matFk} \eeq

Here $\al$, $\be$ and $\ga$ are always real and depend on the parameters $\De$ 
and $D$. The dependence on these parameters is different in the two phases. 
Furthermore, the matrix $F(\vk)$ is Hermitian, since $f(k)=1 + e^{ik} = 
f^*(-k)$. Now, we see that the states with $\vk$ and $-\vk$ are coupled to each 
other and the Hamiltonian is not number conserving, i.e., it has terms of the 
form $a^\dg_k a^\dg_{-k}$. Hence directly diagonalizing $\clyM(\vk)$ will 
not give us the required eigenvalues and eigenvectors. Therefore we use the 
Heisenberg equations of motion for the bosonic operators (see Appendix 
\ref{app.Heisen} for details of the calculation) to obtain the Hamiltonian
\beq h(\vk) ~=~ JS \Big[ \Big( 4\ga \mathbb{1}_{3} - \al F(\vk) \Big) \otimes 
\si_z ~-~ i \be F(\vk) \otimes \si_y \Big]. \label{h23} \eeq
Since $\al$, $\be$ and $\ga$ are different in phases $II$ and $III$, we discuss
the two phases separately below. {(We note here that for Hamiltonians 
which are quadratic in bosonic operators, the standard Bogoliubov treatment 
and the Heisenberg equations of motion give identical energy-momentum 
dispersion relations. For a large number of coupled bosonic operators, we 
find it easier to use the Heisenberg equations of motion since it only 
involves finding the eigenvalues of some matrices, as we have discussed in 
the Appendices.)}

\subsection{Phase $II$}

The DMI term drops out of the equations in this phase and hence the matrix 
$\clyM_{\vk}$ depends only on the anisotropy ratio $\De$. In order to 
obtain the magnon spectrum we use the Heisenberg equations of motion as 
explained in Appendix \ref{app.Heisen}. This method gives us the 
Hamiltonian of the form $h(\vk)$ in Eq.~\eqref{h23} with $\al = (1+\De)/2$, 
$\be = (1-\De)/2$ and $\ga=1$. We thus obtain 
\bea h_{II}(\vk) &=& JS \Big[ \Big(4\mathbb{1}_{3} - \frac{1+\De}{2} F(\vk) 
\Big) \otimes \si_z \non \\
&& ~~~~~~- ~i \frac{1-\De}{2} F(\vk) \otimes \si_y \Big], \label{hk2} \eea
where $F(\vk)$ is the Hermitian matrix from Eq.~\eqref{matFk}.

We find that the spectrum in this phase has no gaps between 
different pairs of bands irrespective of the value 
of $\De$, as shown in the surface plot in Fig.~\ref{2disp}. The top 
band (which is always flat) touches the middle band at 
momentum $\vk = (0,0)$. The middle and bottom bands touch at two points in 
the Brillouin zone given by $\vk = \pm (2\pi/(3\sqrt{3}),-2\pi/3)$. These are 
the same gap closing points as in phase $I$. Finally, the bottom band always 
touches $E=0$ at momentum $\vk = (0,0)$. This corresponds to a Goldstone mode 
in which all the spins are rotated by the same angle while keeping them in the 
$x-y$ plane; as mentioned before, this is a continuous symmetry of the 
Hamiltonian in Eq.~\eqref{Hamiltonian}. We note that the spectrum for 
phase $II$ matches exactly with that of phase $I$ when $\De=1$ and $D=0$, 
i.e., for the isotropic Heisenberg Hamiltonian without any DMI.

Since pairs of bands touch each other at some values of the momentum $\vk$, 
the Berry curvature $\Omega_i (\vk)$ is not well defined at those values of 
$\vk$. Hence the Chern numbers cannot be calculated in any of the bands.

We also calculate the edge state spectrum in this phase by considering an 
infinite strip along the $\hat x$ direction with a finite width of $N_2$ unit 
cells. We solve this by effectively reducing the system to a series 
of one-dimensional 
chains of length $N_2$ each, with the wave function on each chain related to 
the those in the neighboring chains by factors of $e^{\pm ik_1}$. Using this 
along with the Holstein-Primakoff transformations and Heisenberg equations 
of motion, we obtain a $6N_2 \times 6N_2$ spin-wave Hamiltonian which we 
diagonalize to find the spectrum $E$ versus $k_x$. The details of the edge 
state calculation are presented in Appendix \ref{app.edge}.

The edge state spectrum in phase $II$ is shown in Fig.~\ref{2edge}. The 
continuous bands in this figure are projections of the surface plot of the 
bulk spectrum in Fig.~\ref{2disp} on to the $k_x-E$ plane; thus the top band 
in Fig.~\ref{2disp}, which is completely flat, projects on to a single line 
at the top of Fig.~\ref{2edge}. The discrete lines in Fig.~\ref{2edge} show 
the edge state spectra. We see that there are states which are localized near 
the top edge as well as states localized near the bottom edge. 

\begin{widetext}
\begin{center}
\begin{figure}]
\begin{center}
\subfigure[]{\includegraphics[width=7cm]{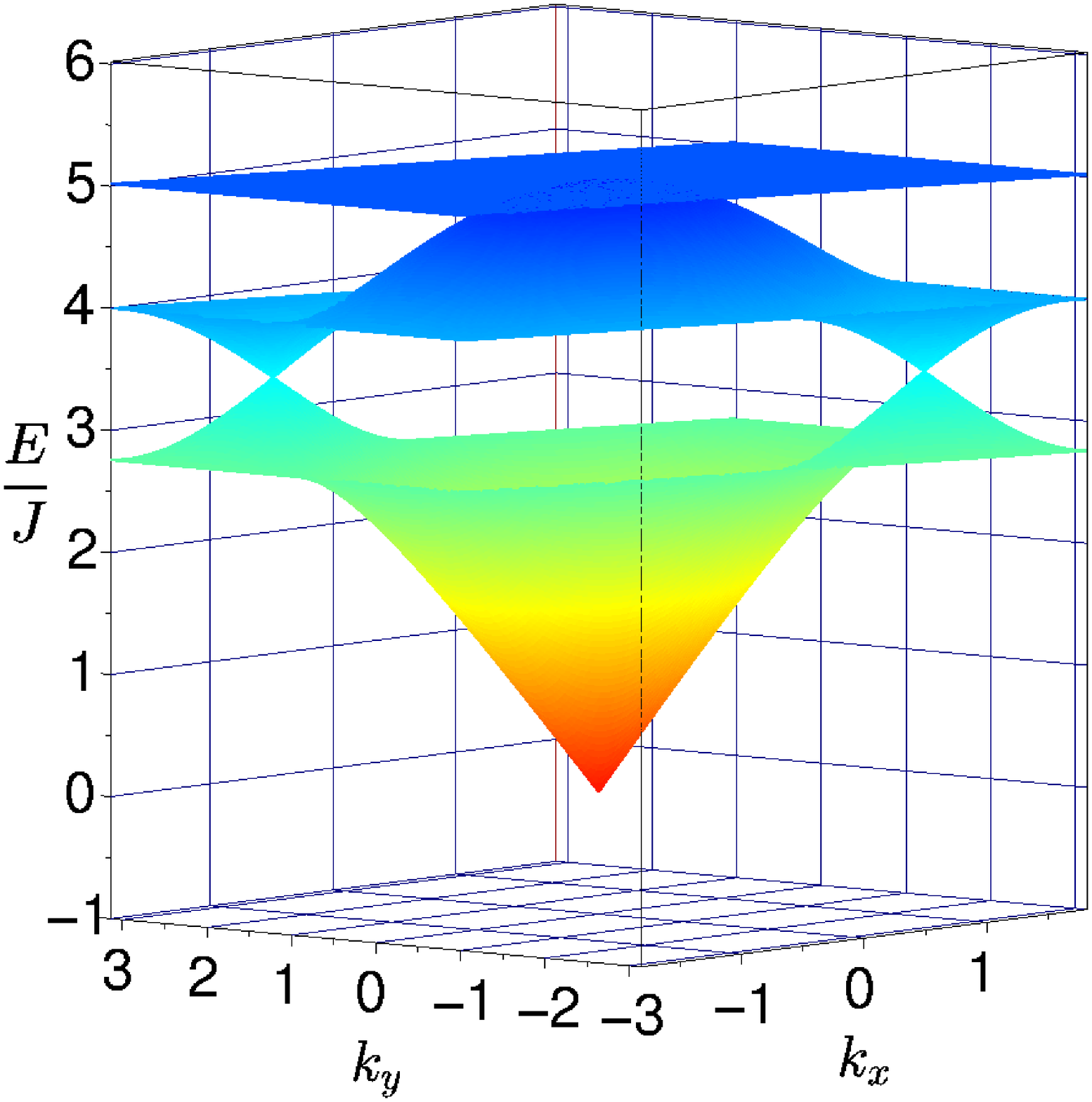}
\label{2disp}}
\subfigure[]{\includegraphics[width=7cm]{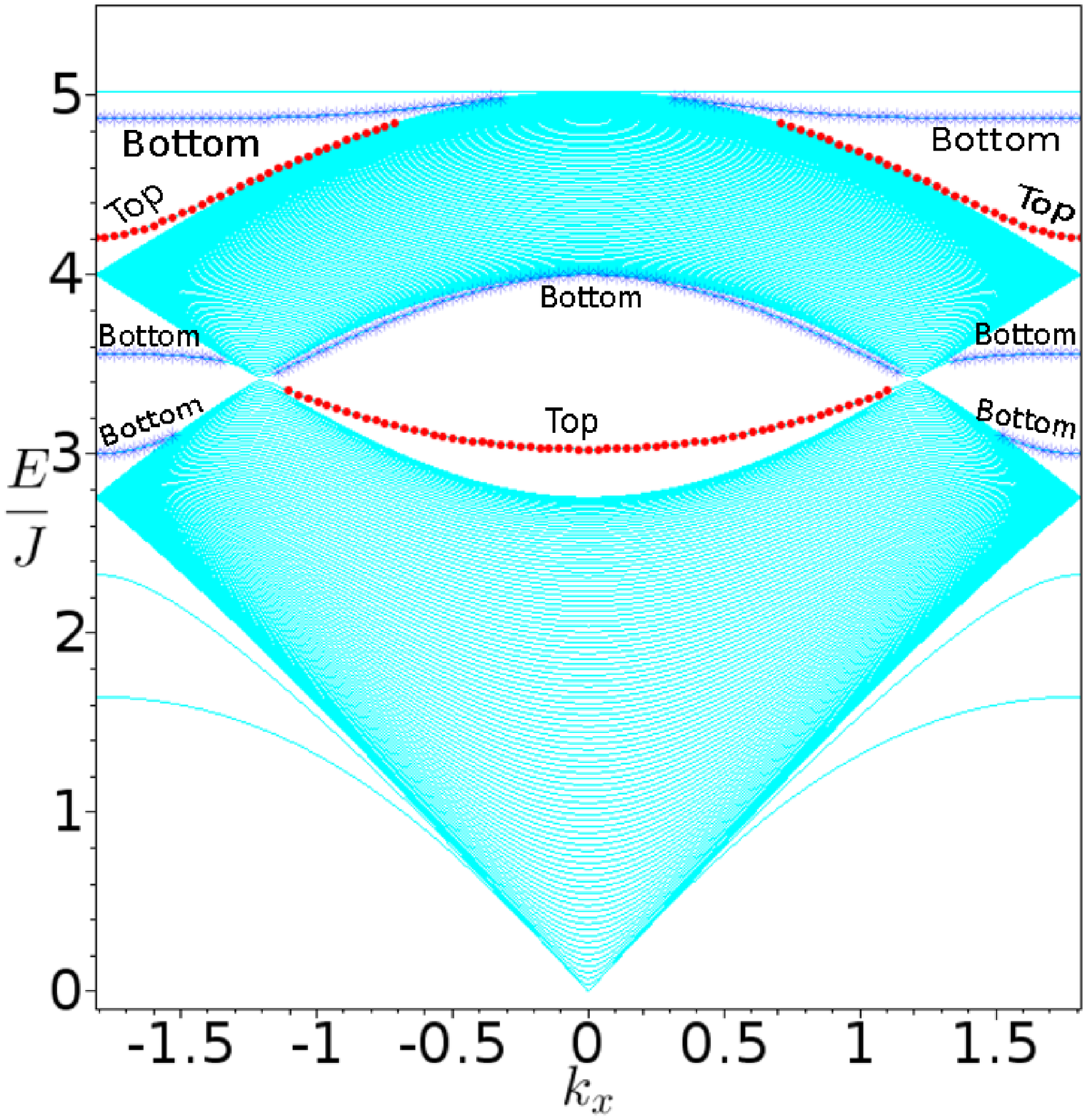}
\label{2edge}} \\
\subfigure[]{\includegraphics[width=7cm]{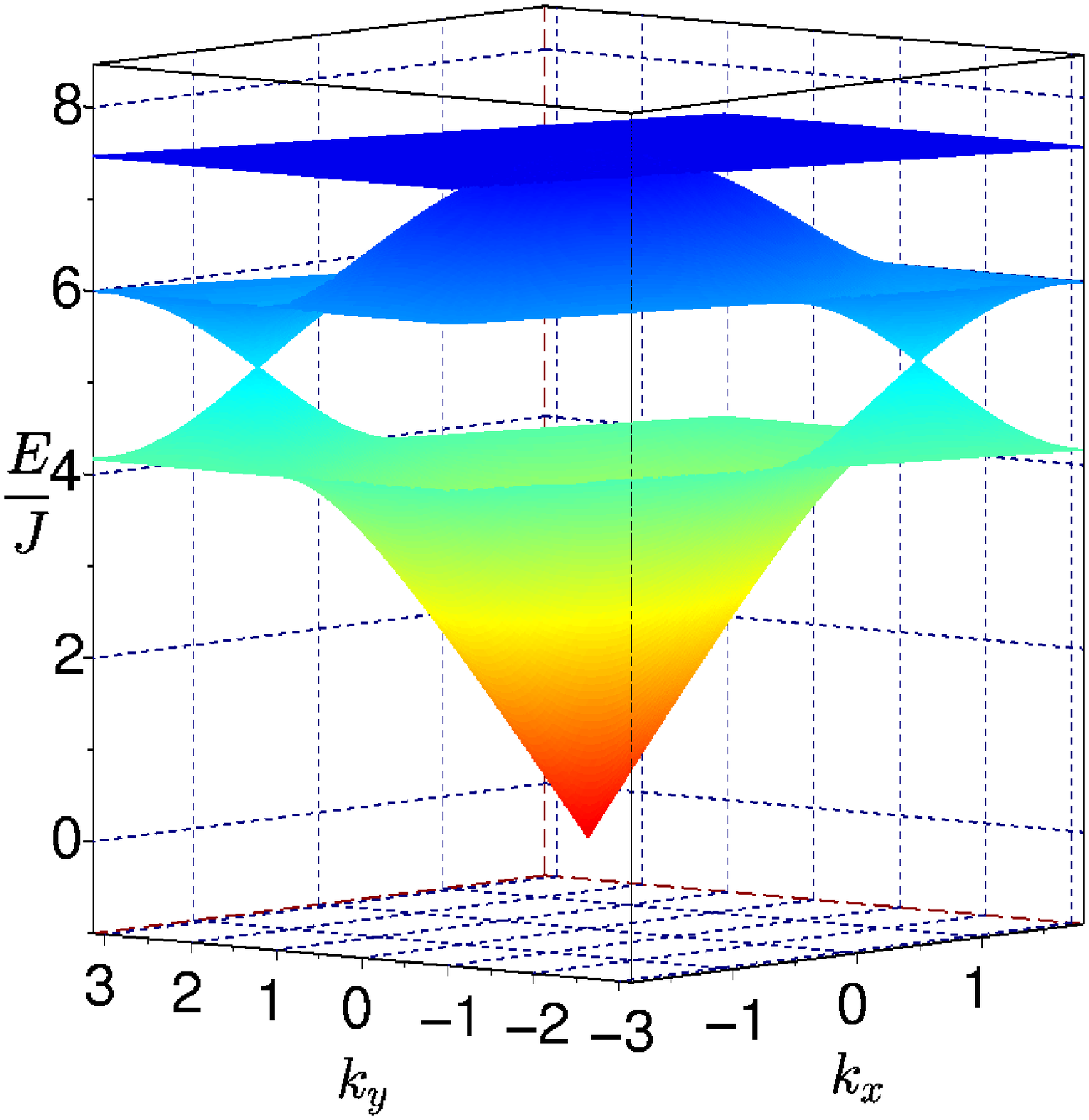}
\label{3disp}}
\subfigure[]{\includegraphics[width=7cm]{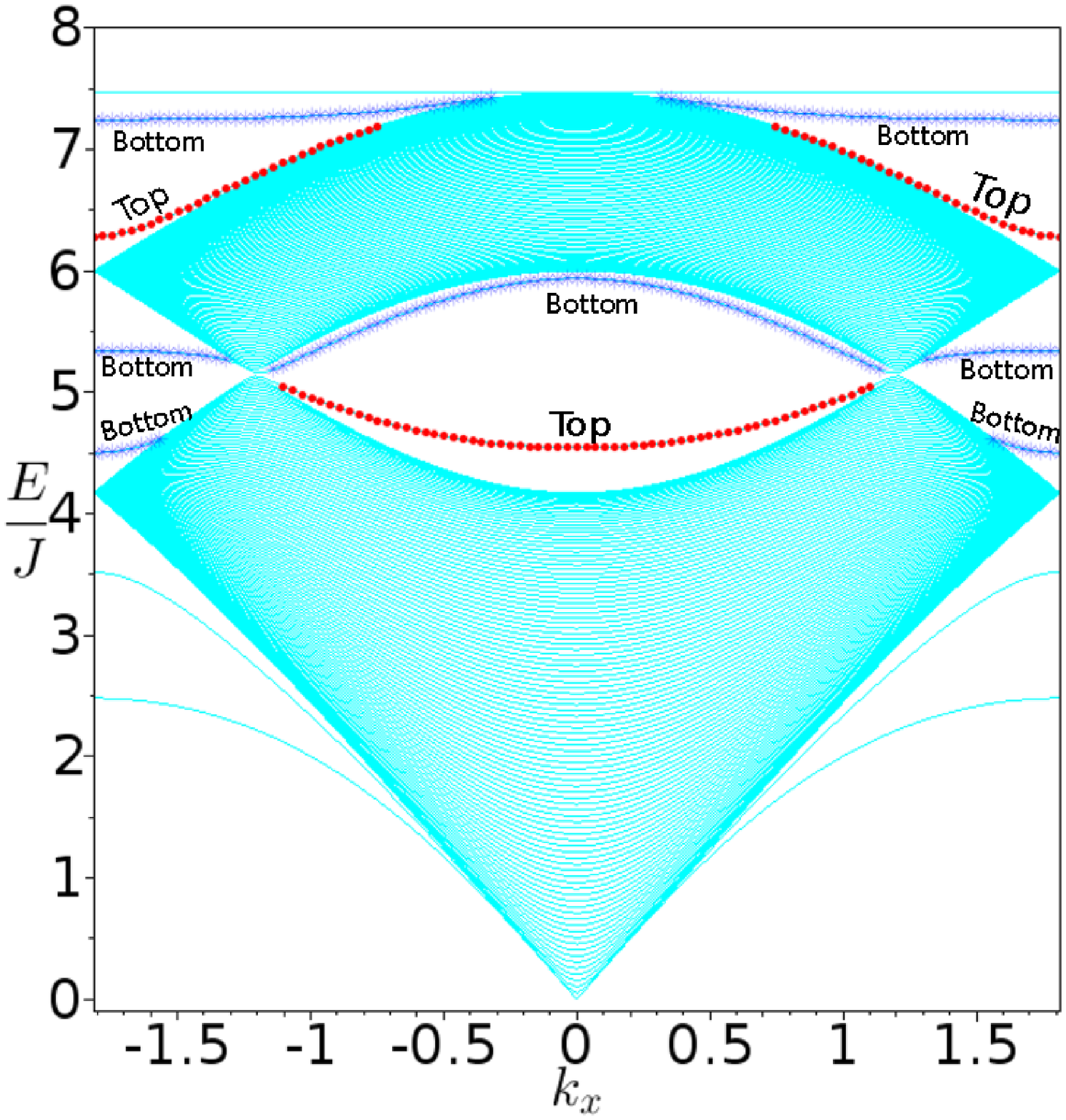}
\label{3edge}}
\caption{Bulk and edge state spectra for phase $II$ (a,b) and phase $III$ 
(c,d). In phase $II$, the magnon spectrum is completely independent of the 
DMI strength; we have set $\De =0.1$. In phase $III$, we set $\De =0.1$ and
$D/J = 4/\sqrt{3}$. The top bands in (a) and (c) are completely flat and hence
appear to be a single line in the corresponding spectra shown in (b) and (d). 
There are no gaps between any pair of bands in either of these phases and 
hence the edge states are not topologically protected. The red (dotted) lines, 
marked as ``Top", show edge states localized near the top edge whereas the blue 
(asterisks) lines, marked as ``Bottom", show edge states at the bottom edge.} 
\end{center}
\end{figure}
\end{center}
\end{widetext}

\subsection{Phase $III$}

In this phase we again follow the same procedure with the Heisenberg equations 
of motion and obtain a Hamiltonian of the same form as in Eq.~\eqref{h23}, 
with $\al = (D'+\De)/2$, $\be = (D'-\De)/2$ and $\ga = D'$, i.e.,
\bea h_{III}(\vk) &=& \Big[ \Big(4D'\mathbb{1}_3-\frac{D'+\De}{2}F(\vk)\Big)
\otimes \si_z \non \\
&& ~~~- ~i\frac{D'-\De}{2}F(\vk)\otimes\si_y \Big], \label{hk3} \eea
where $D' = (\sqrt{3}D/J-1)/2$, and $F(\vk)$ is given in Eq.~\eqref{matFk}.

In contrast to phase $II$, the spin-wave Hamiltonian here depends on both 
$\De$ and the DMI strength $D$. However, interestingly, the qualitative 
results are the same as in phase $II$ as we explain below.

Diagonalizing the Hamiltonian in Eq.~\eqref{hk3} gives the bulk spectrum shown 
in Fig.~\ref{3disp}. In this phase too, as in phase $II$, we find a spectrum 
with no gaps between different pairs of bands, and a dispersionless top band. 
The top and middle bands touch at $\vk = (0,0)$, the middle and bottom bands 
touch at $\vk = \pm (2\pi/(3\sqrt{3}), -2\pi/3)$, and the bottom band touches 
$E=0$ at $\vk = (0,0)$ corresponding to the Goldstone mode discussed above. 
The gapless nature of the spectrum results in ill-defined Chern numbers; hence 
the edge states in this phase are not topologically protected. 
Figure~\ref{3edge} shows the bulk and edge state spectrum in phase $III$.

\section{Discussion}
\label{sec:discussion}

In this paper we have studied a spin model on the kagome lattice with $XXZ$ and 
Dzyaloshinskii-Moriya interactions between nearest-neighbor sites. In the limit 
of large spin $S$, a classical analysis shows that this system has five 
different phases with different ground-state spin configurations. We then use
the Holstein-Primakoff transformation from spins to bosons in order to go to 
the next order in a $1/S$ expansion; this gives the energy dispersion of the
magnons in the bulk. 

We find that one of the phases, called phase $I$, is
rather simple; in the ground state, all the spins point in the $+ \hat z$ or 
$- \hat z$ direction. We find that this phase consists of four distinct 
topological phases in which the bulk bands have different values of the Chern 
numbers. We have studied strips of the system which are infinitely long in one 
direction and have a finite width in the other direction; such a system hosts 
states whose wave functions are localized near one of the two edges and whose 
energies lie within the gaps of the bulk bands. The number of edge states is 
related to the Chern numbers which confirms that this phase is topological. We
calculate the thermal Hall conductivity and find that this too can distinguish 
between the different topological phases lying within phase $I$. All these 
results are valid for any finite value of $S$, which includes spin-1/2. 
This is an interesting model in which a phase
with a particular ground-state spin configuration contains multiple phases in 
which the magnon bands have different topological structures. Thus the ground
state does not uniquely determine the topological phase of the magnons.

We have used spin-wave theory to study three of the other phases 
($II - IV$). In these phases, we find that there are no gaps between pairs 
of bulk bands; hence the Chern numbers of the bands cannot be calculated, and 
these phases are not topological. A long strip of the system again has edge 
states in these phases but these are not topologically protected.

In the classical limit ($S \to \infty$), the model studied in this paper
is ordered at all points in the space given by $(\Delta, D)$. Namely, for
all values of $(\Delta, D)$, the model has a particular classical ground 
state or a set of ground states which are related to each other by rotations 
in the $x-y$ plane. However, the situation may be different for any finite 
value of $S$. For finite $S$, it is not clear if we will 
have a ground state with long range order which breaks the $U(1)$ symmetry 
of rotations in the $x-y$ plane. However, in phase I, the classical ground 
state, which consists of all the spins pointing in the $\hat z$ direction 
(note that this state does not break the $U(1)$ rotational symmetry), is also 
the ground state for any finite value of $S$, and this ground state is an 
exact eigenstate of the Hamiltonian.

The model studied in this paper may be experimentally realizable in 
systems of ultracold atoms trapped in optical lattices. It is known that such 
systems can give rise to effective spin models with anisotropic $XXZ$ 
interactions~\cite{struck} and with Dzyaloshinskii-Moriya 
interactions~\cite{zhang,gong}. It should therefore be possible in the future 
to realize systems with both $XXZ$ and Dzyaloshinskii-Moriya interactions.

We end by mentioning some possible directions for future investigations. 
First, it would be interesting to study the behavior of the system in more 
detail in phase $V$. In this phase, since the spin configuration on the three 
sublattices is generally not coplanar in the classical ground state, we can 
expect a more complicated magnon spectrum and edge state structure. Second, 
in the colinear phases ($I - II$) and coplanar phases ($III - IV$), adding a 
next-nearest-neighbor interaction can significantly alter the magnon spectrum 
as well as the topological characters of the bands, thus affecting the 
behavior of the edge states~\cite{mook1}. 
The combined effects of anisotropy, the next-nearest-neighbor coupling 
and the Dzyaloshinskii-Moriya interaction may result in an even richer phase 
diagram. Third, topological phases of spin systems with Dzyaloshinskii-Moriya 
interactions on other kinds of lattices are also of interest~\cite{kim}.
Finally, it would be interesting to investigate the effect of an 
external magnetic field. A field applied along the ${\hat z}$ direction
would keep the ground-state spin configuration in phase $I$ as it is, although
it would change the ground-state energy. This is because phase $I$ already 
has all the spins pointing along the $\hat z$ direction, and the Zeeman term 
for a field pointing along that direction commutes with the Hamiltonian.
Hence the ground state and one-magnon states that we have studied in this
phase are exact eigenstates of the Hamiltonian for all values of $S$;
the Zeeman term simply shifts the energies of all the one-magnon states by
the same amount. Our analysis of the topological phases would therefore 
remain unchanged in the presence of such a field. However a field in the 
$\hat z$ direction would significantly alter the phases 
$II-IV$ since the ground state in those phases have the spins pointing in the 
$x-y$ plane. On the other hand, applying a magnetic field along the 
$\hat x$ direction, keeps phase $II$ the same but changes the ground-state 
configurations in all the other phases. This implies that an applied magnetic 
field can also alter the phase boundaries separating the five main phases
as well as the four topological phases within phase $I$.

\vspace*{.5cm} 
\section*{Acknowledgments}

We thank H. Katsura for useful comments. D.S. thanks the Department of Science
and Technology, India for Project No. SR/S2/JCB-44/2010 for financial support.

\appendix

\section{Holstein-Primakoff transformation in phase $III$}
\label{app.HPIII}

In this phase the ground-state spins point in the directions of the vectors 
$-\hn_2$, $\hn_1 - \hn_2$ and $\hat x$ (as shown in Fig.~\ref{classical}) 
for sites in the $A$, $B$ and $C$ sublattices respectively. The directions 
$\hn_1$ and $\hn_2$ are along the $A-B$ and $A-C$ bonds of the unit cell. 
However, the Hamiltonian in Eq.~\eqref{Hamiltonian} is written in terms of 
the spins in the $\hat x$ and $\hat y$ 
directions. In order to use the Holstein-Primakoff transformations in 
Eqs.~\eqref{HP23} and \eqref{HP3}, we must therefore write the Hamiltonian 
in terms of this new basis. If $R(\theta)$ is the rotation matrix in the
$x-y$ plane, i.e.,
\beq \mathcal{R}(\theta) ~=~ \begin{pmatrix} \cos\theta & -\sin\theta \\ 
\sin\theta &\cos\theta \end{pmatrix}, \eeq
and $\hn_3 = \hn_1 - \hn_2$ and $\hn'$ denote the directions orthogonal to 
$\hn$ such that $\hn \times \hn'$ is along $\hat z$, then the $x$ and $y$ 
components of the spin at each site are related to the new basis as
\beq \begin{pmatrix}A^x \\ \\ A^y\end{pmatrix} ~=~ \mathcal{R} 
(\frac{2\pi}{3}) \begin{pmatrix}A^{-\hn_2} \\ \\ 
A^{-\hn'_2} \end{pmatrix}, \eeq
and
\beq \begin{pmatrix} B^x \\ \\ B^y \end{pmatrix} ~=~ \mathcal{R} (-
\frac{2\pi}{3}) \begin{pmatrix} B^{\hn_3} \\ \\ B^{\hn'_3} \end{pmatrix}. \eeq
Since we have assumed the spins on the $C$ sublattice to be aligned along the 
$+\hat x$ direction, we need not transform those. Therefore, the $x-y$ part of 
$H_H$ in Eq.~\eqref{Hamiltonian}, i.e., 
\bea H^{xy}_{H} &=& - J S \displaystyle \sum_{\substack{\la\vn\vn'\ra \\ 
\al=x,y}} (A^{\al}_{\vn} B^{\al}_{\vn'}+B^{\al}_{\vn}C^{\al}_{\vn'}+
C^{\al}_{\vn}A^{\al}_{\vn'}) \eea
can be written in terms of the new basis as
\bea H^{xy}_{H}= &-& J S \displaystyle\sum_{{\la \vn \vn' \ra}} {\Bigg\{}
\begin{pmatrix} A^{-\hn_2}_\vn & A^{-\hn'_2}_\vn\end{pmatrix}
\mathcal{R} (-\frac{2\pi}{3})
\begin{pmatrix} B^{\hn_3}_{\vn'} \\ \\ B^{\hn'_3}_{\vn'} \end{pmatrix} \non \\
&+&\begin{pmatrix} B^{\hn_3}_\vn & & B^{\hn'_3}_\vn \end{pmatrix} 
\mathcal{R} (-\frac{2\pi}{3})
\begin{pmatrix}C^x_{\vn'} \\ \\ C^y_{\vn'} \end{pmatrix}\non\\
&+& \begin{pmatrix} C^{x}_\vn & & C^{y}_\vn \end{pmatrix}
\mathcal{R} (-\frac{2\pi}{3}) \begin{pmatrix} A^{-\hn_2}_{\vn'} \\ \\ 
A^{-\hn'_2}_{\vn'} \end{pmatrix}{\Bigg\}}, \eea
where nearest neighbors are denoted by $\la \vn, \vn' \ra$. 
The $z-$part of $H_H$ remains the same as in phase $I$, i.e.,
\beq H^z_H = -J S \De \displaystyle\sum_{{\la \vn \vn' \ra}} (A^{z}_{\vn}
B^{z}_{\vn'}+B^{z}_{\vn}C^{z}_{\vn'}+C^{z}_{\vn}A^{z}_{\vn'}). \eeq
For the DMI term $H_{DM}$, we write terms of the type $A^x B^y-A^y B^x$ in the
new basis, which gives us
\bea H_{DM} = &DS& \displaystyle\sum_{{\la \vn \vn' \ra}}{\Bigg\{}
\begin{pmatrix} A^{-\hn_2}_\vn & & A^{-\hn'_2}_\vn\end{pmatrix}
~\mathcal{R} (\frac{\pi}{6})
\begin{pmatrix} B^{\hn_3}_{\vn'} \\ \\ B^{\hn'_3}_{\vn'} \end{pmatrix} \non \\
&+&\begin{pmatrix} B^{\hn_3}_\vn & & B^{\hn'_3}_\vn \end{pmatrix} 
~\mathcal{R} (\frac{\pi}{6})
\begin{pmatrix}C^x_{\vn'} \\ \\ C^y_{\vn'} \end{pmatrix}\non\\
&+&\begin{pmatrix} C^{x}_\vn & & 
C^{y}_\vn \end{pmatrix}~\mathcal{R} (\frac{\pi}{6}) \begin{pmatrix} 
A^{-\hn_2}_{\vn'} \\ \\ A^{-\hn'_2}_{\vn'} \end{pmatrix}{\Bigg\}}. 
\label{HDM_new} \eea
Using the Holstein-Primakoff transformations in Eqs.~\eqref{HP23} and 
\eqref{HP3} and ignoring all terms higher than second order in the bosonic 
creation and annihilation operators, we find that $H_{DM} \simeq -D
\sqrt{3}H^{xy}_H$. Therefore, in phase $III$ we get a very simple form of 
the spin-wave Hamiltonian,
\beq H_{III} ~\simeq~ (1-D\sqrt{3})H^{xy}_H ~+~ H^z_{H}. \eeq
We then use the Fourier transform of the bosonic operators from Eq. \eqref{FT}, 
to obtain the following Hamiltonian in terms of the momentum $\vk$,
\bea H(\vk) ~=~ \displaystyle \sum_{\vk} ~\Psi^\dg (\vk) ~\clyM(\vk) 
~\Psi^(\vk), \label{hvk2} \eea
such that
\beq \Psi^\dg (\vk) ~=~ \begin{pmatrix}a^\dg_\vk & b^\dg_\vk & c^\dg_\vk & 
a_{-\vk} & b_{-\vk} & c_{-\vk} \end{pmatrix}, \eeq
and $M(\vk)$ is a $6 \times 6$ matrix
\bea \clyM(\vk) &=& JS \Big[ \Big(4D' \mathbb{1}_3 -\frac{D'+\De}{2} F(\vk)
\Big) \otimes \mathbb{1}_2 \non \\
&& ~~~~~~- ~\frac{D'-\De}{2}F(\vk)\otimes \si_x \Big], \eea
where $D' = (\sqrt{3}D/J-1)/2$, and $F(\vk)$ is given in Eq.~\eqref{matFk}.
After this, we use the Heisenberg equations of motion as shown in Appendix 
\ref{app.Heisen} for further analysis.

\section{Energy spectrum using Heisenberg equations of motion} 
\label{app.Heisen} 

Since the Hamiltonian in Eq.~\eqref{hvk2} mixes the $+\vk$ and $-\vk$ states, 
we use the following Heisenberg equations of motion to calculate the magnon
spectrum in phases $II$ and $III$. 
\bea \Big[a^{\dg}_{\vk},H(\vk)\Big] &=& i\dot{a}^{\dg}_{\vk} ~=~ E 
a^{\dg}_{\vk}, \non \\
\Big[b^{\dg}_{\vk},H(\vk)\Big] &=& i\dot{b}^{\dg}_{\vk} ~=~ E 
b^{\dg}_{\vk}, \non \\
\Big[c^{\dg}_{\vk},H(\vk)\Big] &=& i\dot{c}^{\dg}_{\vk} ~=~ E 
c^{\dg}_{\vk}, \label{comm} \eea
where we have assumed that all the bosonic operators evolve in time as 
$e^{-iEt }$. Using $H(\vk)$ from Eq.~\eqref{ham23} to calculate the
commutators in Eq.~\eqref{comm}, we obtain the following two matrix equations, 
\begin{scriptsize}
\bea \begin{pmatrix}
~M_{11} & ~M_{12} & ~M_{13} & ~M_{14} & ~M_{15} & ~M_{16}\\ \\
~M_{21} & ~M_{22} & ~M_{23} & ~M_{24} & ~M_{25} & ~M_{26}\\ \\
~M_{31} & ~M_{32} & ~M_{33} & ~M_{34} & ~M_{35} & ~M_{36}\\ \\
-M_{41} & -M_{42} & -M_{43} & -M_{44} & -M_{45} & -M_{46}\\ \\
-M_{51} & -M_{52} & -M_{53} & -M_{54} & -M_{55} & -M_{56}\\ \\
-M_{61} & -M_{62} & -M_{63} & -M_{64} & -M_{65} & -M_{66}\\ \\
\end{pmatrix}\begin{pmatrix}a_\vk \\ \\b_\vk \\ \\c_\vk \\ \\a^\dg_{-\vk} \\\\
b^\dg_{-\vk} \\ \\c^\dg_{-\vk} \end{pmatrix} = E \begin{pmatrix}a_\vk \\ \\
b_\vk \\ \\ c_\vk \\ \\ a^\dg_{-\vk} \\ \\
b^\dg_{-\vk} \\ \\ c^\dg_{-\vk} \end{pmatrix} \non \\
&& \label{com1} \eea
\end{scriptsize}
and
\begin{scriptsize}
\bea \begin{pmatrix}
-M_{11} & -M_{21} & -M_{31} & -M_{41} & -M_{51} & -M_{61}\\ \\
-M_{12} & -M_{22} & -M_{32} & -M_{42} & -M_{52} & -M_{62}\\ \\
-M_{13} & -M_{23} & -M_{33} & -M_{43} & -M_{53} & -M_{63}\\ \\
~M_{14} & ~M_{24} & ~M_{34} & ~M_{44} & ~M_{54} & ~M_{64}\\ \\
~M_{15} & ~M_{25} & ~M_{35} & ~M_{45} & ~M_{55} & ~M_{65}\\ \\
~M_{16} & ~M_{26} & ~M_{36} & ~M_{46} & ~M_{56} & ~M_{66}\\ \\
\end{pmatrix}\begin{pmatrix}a^\dg_\vk \\ \\ b^\dg_\vk \\ \\ c^\dg_\vk \\ \\ 
a_{-\vk} \\ \\ b_{-\vk} \\ \\ c_{-\vk} \end{pmatrix} = E \begin{pmatrix}
a^\dg_\vk \\ \\ b^\dg_\vk \\ \\ c^\dg_\vk \\ \\ 
a_{-\vk} \\ \\
b_{-\vk} \\ \\ c_{-\vk} \end{pmatrix}, \non \\
&& \label{com2} \eea 
\end{scriptsize}
which we can write more compactly as $h \Psi(k_1) = E \Psi(k_1)$ and 
$\tilde{h} \tilde{\Psi}(k_1) = \tilde{E}\tilde{\Psi}(k_1)$, where 
$\tilde{\Psi} = (\Psi^\dag)^T$.

It turns out that the matrix $\clyM(\vk)$, with matrix elements given by
$M_{ij}$, can be written as
\beq \clyM(\vk) ~=~ JS \Big[ \Big( 4 \ga \mathbb{1}_{3} - \al F(\vk) \Big) 
\otimes \mathbb{1}_{2}- \be F(\vk) \otimes \si_x \Big], \eeq
where $F(\vk)$ is given in Eq.~\eqref{matFk}, and $\al$, $\be$ and $\ga$ 
(which are all real-valued functions of $\De$ and $D$) take different values 
in phases $II$ and $III$. In block form, we have
\beq \clyM(\vk) ~=~ JS \begin{pmatrix} 4\ga \mathbb{1}_3 -\al F(\vk) & 
-\be F(\vk) \\ \\ 
-\be F(\vk) & 4\ga\mathbb{1}_3 -\al F(\vk) \end{pmatrix}. \eeq
This means that the matrix appearing on the left hand side of
Eq.~\eqref{com1} is given by
\beq h(\vk) ~=~ JS \begin{pmatrix} 4\ga\mathbb{1}_3-\al F(\vk)& -\be 
F(\vk) \\ \\ \be F(\vk) &-(4\ga \mathbb{1}_3-\al F(\vk)) \end{pmatrix}. \eeq
In a more compact form this is the same as in Eq. \eqref{h23}, i.e.,
\beq h(\vk) ~=~ JS \Big[ \Big( 4\ga\mathbb{1}_{3} - \al F(\vk) \Big) \otimes 
\si_z - i \be F(\vk) \otimes \si_y \Big], \eeq
whose eigenvalues give the magnon spectra in phases $II$ and $III$.
Since $F(\vk) = F^\dg(\vk)$ and $\al$ and $\be$ are real, we have $h = 
\tilde{h}^\dg$. Hence, diagonalizing one of these, say, $h(\vk)$ is sufficient 
to obtain the magnon spectrum. 

In phase $II$, $\al = (1+\De)/2$, $\be = (1-\De)/2$ and $\ga=1$ which gives us
Eq.~\eqref{hk2}. The magnon spectrum in this phase is completely independent 
of the DMI strength $D$.
On the other hand, in phase $III$, $\al=(D'+\De)/2$, $\be=(D'-\De)/2$ and 
$\ga = D'$ where $D'=(\sqrt{3}D/J-1)/2 $ which gives us Eq.~\eqref{hk3}.

\section{Calculating the edge states}
\label{app.edge}

We consider a strip which is $N_2$ unit cells wide in the $\hn_2$ direction 
and is infinitely long along the $\hat x$ direction ($\hn_1$ direction). As 
the system is translationally invariant along $\hn_1$, $k_1$ is a good quantum
number; the system can be reduced to a single chain which has $N_2$ unit
cells (i.e., $3N_2$ sites) with wave functions which are related to those of
neighboring chains by factors of $e^{\pm ik_1}$. We Fourier transform 
the bosonic operators only in the $\hn_1$ direction
\beq (a,b,c)_{\vn} ~=~ \displaystyle \sum_{k_1} (a,b,c)_{(k_1,n_2)} ~
e^{ik_1 n_1}. \eeq
We then use suitable Holstein-Primakoff transformations depending upon 
the phase we are working in, and we retain terms only up to second order
in the bosonic operators as usual. 

\subsection{Phase $I$}

In phase $I$, the procedure described above gives us a number conserving 
$3N_2 \times 3N_2$ Hamiltonian which is a function of the momentum $k_1$, i.e.,
\bea H^{1d}_{I} ~=~ \displaystyle \sum_{k_1} ~{\Psi^\dg}(k_1) ~h(k_1) ~
{\Psi}(k_1), \eea
where
\bea {\Psi^\dg}_{k_1} = \begin{pmatrix}a^\dg_{k_1,1} & b^\dg_{k_1,1} & 
c^\dg_{k_1,1} & \cdots & a^\dg_{k_1,N_2} & b^\dg_{k_1,N_2} & 
c^\dg_{k_1,N_2}&\end{pmatrix}. \non \\ \eea
Here the site index $n_2$ increases from bottom to top along the direction 
$\hn_2$, i.e., the sites on the lower edge are labeled as $n_2=1$ while on 
the top edge they are labeled as $n_2=N_2$. Diagonalizing $h(k_1)$ gives us 
the energy levels and wave functions of the bulk as well as edge states as 
functions of momentum $k_1$.

\subsection{Phases $II$ and $III$}

In this case we get a non-number conserving Hamiltonian which is a $6N_2 \times 
6N_2$ matrix that mixes the $+k_1$ and $-k_1$ states, i.e.,
\bea H^{1d} ~=~ \displaystyle \sum_{k_1} ~{\Psi^\dg}(k_1) ~\clyM^{1d}(k_1)~
{\Psi}(k_1), \eea
where the $6N_2$ component row vector $\Psi^\dg (k_1)$ is 
\bea && {\Psi^\dg}(k_1) ~=~ {\Big(}a^\dg_{k_1,1}~~b^\dg_{k_1,1}~~
c^\dg_{k_1,1} ~~ a_{-k_1,1} ~~b_{-k_1,1}~~c_{-k_1,1} \non \\ 
&& \cdots ~~a^\dg_{k_1,N_2}~~b^\dg_{k_1,N_2}~~c^\dg_{k_1,N_2}~~
a_{-k_1,N_2} ~~b_{-k_1,N_2}~~c_{-k_1,N_2} {\Big)}. \non \\ \eea
We again use the Heisenberg equations of motion for the bosonic operators 
at each site of the chain along the $\hn_2$ direction,
\bea \Big[a^{\dg}_{\pm k_1,n_2},H^{1d}(k_1)\Big] &=& i{\dot{a}^{\dg}_{\pm
k_1,n_2}} ~=~ E a^{\dg}_{\pm k_1,n_2}, \non \\
\Big[b^{\dg}_{\pm k_1,n_2},H^{1d}(k_1)\Big] &=& i{\dot{b}^{\dg}_{\pm 
k_1,n_2}} ~=~ E b^{\dg}_{\pm k_1,n_2}, \non \\
\Big[c^{\dg}_{\pm k_1,n_2},H^{1d}(k_1)\Big] &=& i{\dot{c}^{\dg}_{\pm 
k_1,n_2}} ~=~ E c^{\dg}_{\pm k_1,n_2} \label{comm1d} \eea
to obtain the spectrum $E$ versus $k_x$. This gives us two sets of matrix 
equations similar to Eqs.~\eqref{com1} and \eqref{com2}, except that now we 
have $6N_2$ components instead of just six, i.e., $h^{1d}(k_1) \Psi(k_1) 
= E \Psi(k_1)$ and $\tilde{h}^{1d}(k_1) \tilde{\Psi}(k_1) = \tilde{E}
\tilde{\Psi}(k_1)$,
where $\tilde\Psi=(\Psi^\dg)^T$, and $h^{1d}$ and $\tilde{h}^{1d}$ are related 
to each other by Hermitian conjugation. Diagonalizing either one of these $6N_2
\times 6N_2$ matrices (either $h^{1d}(k_1) $ or $\tilde{h}^{1d}(k_1)$) gives 
us the spectrum $E$ versus $k_1$, and hence $E$ versus $k_x$, since
$k_x=k_1/\sqrt{3}$.

\end{document}